# Pulsating propagation and extinction of hydrogen detonations in ultrafine water sprays

Yong Xu, Huangwei Zhang*

*Department of Mechanical Engineering, National University of Singapore, 9 Engineering Drive 1, Singapore, 117576, Republic of Singapore*

## Abstract

The Eulerian-Lagrangian method is applied to simulate pulsating propagation and extinction of stoichiometric hydrogen/oxygen/argon detonations in ultrafine water sprays. The emphasis is laid on characterization of the unsteady phenomena and evolutions of chemical / flow structures in pulsating propagation and extinction. Three detonation propagation modes are found: (1) pulsating propagation, (2) propagation followed by extinction, and (3) immediate extinction. For pulsating detonation, within one cycle, the propagation speeds and the distance between reaction front (RF) and shock front (SF) change periodically. The pulsating phenomenon originates from the interactions between gas dynamics, chemical kinetics, and droplet dynamics inside the induction zone. Multiple pressure waves are emanated from the RF within one cycle, which overtake and intensify the lead SF. An autoigniting spot arises in the shocked gas after the contact surface. The relative locations of SF, RF, shock-frame sonic point, and two-phase contact surface remain unchanged in a pulsating cycle, but their distances have periodic variations. Moreover, the unsteady behaviors of detonation extinction include continuously increased distance between the RF and SF and quickly reduced pressure peaks, temperature, and combustion heat release. The decoupling of RF and SF leads to significantly increasing chemical timescale of the shocked mixture. The hydrodynamic structure also changes considerably when detonation extinction occurs. Moreover, the predicted map of detonation propagation and extinction illustrates that the critical mass loading for detonation extinction reduces significantly when the droplet size becomes smaller. It is also found that for the same droplet size, the average detonation speed monotonically decreases with water mass loading. However, detonation speed and pulsating frequency have a non-monotonic dependence on droplet size under a constant water mass loading.

* Corresponding author. Tel.: +65 6516 2557; Fax: +65 6779 1459.
*E-mail address*: huangwei.zhang@nus.edu.sg.

## 1. Introduction

There are growing concerns about carbon dioxide ($CO_2$) emissions from fossil fuel combustion and utilization of low-carbon fuels is in high demand. Hydrogen ($H_2$) is deemed a clean fuel because its oxidation does not generate $CO_2$. However, compared to conventional hydrocarbons, $H_2$ is more chemically reactive, with lower ignition energy and wider flammability limit [1]. Therefore, inhibition of hydrogen ignition and explosion is of utmost importance for its practical applications. Water spray is an ideal mitigant for gas explosions [2,3], because it can absorb heat from gas phase due to large heat capacity and latent heat of evaporation [4]. Besides, the sprayed water droplets have large specific surface area and low terminal velocity, and hence can circulate in explosion area in a manner of a total flooding gas.

There have been numerous studies available about the mechanisms of water sprays in inhibiting flammable gas detonation. For instance, Thomas et al. [5] attribute detonation extinction to high heat loss due to water droplets. The water droplet diameter and loading densities are identified as key factors for detonation inhibition. Niedzielska et al. [6] also observe that small droplets have strong influence on detonation quenching due to their fast evaporation rate. Jarsalé et al. [7] find that water droplets do not alter the ratio of the hydrodynamic thickness to the cell size, but it can affect detonation stability. Besides the foregoing experimental work, Song and Zhang [8] simulate methane detonation with water mists and find that the water droplets considerably reduce the flame temperature. Recently, Watanabe et al. [9] observe that the numerical soot foils of dilute water spray detonation become more regular compared to the droplet-free detonations, and the hydrodynamic thickness is altered by the direct interactions between detonation wave and water droplets. Their results also show that droplet breakup mainly occurs near the shock front, and the average diameter of the disintegrated water droplets is independent on the initial propagation velocity of the shock front [10]. However, unsteady detonation behaviors caused by the added water droplets have not been focused on in these studies.

Pulsating propagation is one of the important unsteady phenomena in detonative combustion, because of the interactions between gas dynamics and chemical kinetics [11]. The wave-interaction theory by Toong and his co-workers [12,13] indicates that pulsating instability results from the periodic interactions between shock front and wave structures due to the formation of new reaction zone. This theory has been confirmed in Refs. [14–17]. Furthermore, Short et al. [18] demonstrate that the instability of regular detonation oscillations is induced by compression and expansion waves within the induction zone. For irregular oscillations, it is driven by the decoupling / coupling of reaction front and shock front. Sussman [19] find that the ratio of the heat release time to the induction time affects the longitudinal oscillation mode. Ng et al. [20,21] use bifurcation theory to analyse detonation instability and propose a stability parameter depending on the induction and reaction zone length.

Recently, Han et al. [22,23] identify four pulsating detonation modes, including mildly unstable detonation with multi- and single-period pulsations. Kim et al. [24] reveal that nonlinear temporal patterns is a function of the perturbation wavelength. This finding demonstrates that one-dimensional gaseous chaotic pulsating oscillations can be regularized by the coupling between the intrinsic instability and small perturbation. Zhao et al. [25] show that pulsating detonation of *n*-heptane/air mixtures occurs under off-stoichiometric conditions. They also observe that the pulsating frequency is considerably influenced by equivalence ratio, initial pressure, and temperature. Nonetheless, to the best of our knowledge, no studies are available hitherto about how the dispersed phase affects the pulsating detonation propagation.

In this work, Eulerian–Lagrangian method with two-way coupling is employed to predict incident detonations in ultrafine water sprays. A one-dimensional configuration filled with hydrogen/oxygen/argon mixture and ultrafine water droplets is considered, and the effects of water droplet properties on incident detonations are studied in detail. Our work will be focused on the following aspects of gas–liquid two-phase detonations: (1) identification and characterization of detonation pulsating propagation and extinction in gas-liquid two-phase mixtures; (2) evolutions of the hydrodynamic structure of unsteady spray detonations; and (3) quantification of chemical reaction information with chemical explosive mode analysis [26–29]. This work is devoted to understanding fundamental detonation phenomenon in two-phase medium and providing possible knowledge for practical hydrogen safe utilization. The manuscript is structured as below. The mathematical model will be clarified in Section 2, whilst the physical model will be detailed in Section 3. The results and discussion will be presented in Section 4, followed by main conclusions in Section 5. Sensitivity analysis of our numerical implementations can be found in supplementary document.

## 2. Mathematical model

The Eulerian−Lagrangian method is used to simulate gas−droplet two-phase reactive flows. The Eulerian equations for gas phase and Lagrangian equations for droplet phase are solved by a two-phase compressible flow solver, *RYrhoCentralFoam* [30,31], developed from OpenFOAM 6.0 [32]. The formulation and numerical method are presented below.

### 2.1 Gas phase

The governing equations of mass, momentum, energy, and species mass fraction are solved for gas phase. They respectively read

$$\frac{\partial \rho}{\partial t} + \nabla \cdot [\rho \mathbf{u}] = S_{mass}, \tag{1}$$

$$\frac{\partial (\rho \mathbf{u})}{\partial t} + \nabla \cdot [\mathbf{u}(\rho \mathbf{u})] + \nabla p + \nabla \cdot \mathbf{T} = \mathbf{S}_{mom}, \tag{2}$$

$$\frac{\partial (\rho \boldsymbol{E})}{\partial t} + \nabla \cdot [\mathbf{u}(\rho \boldsymbol{E} + \boldsymbol{p})] + \nabla \cdot [\mathbf{T} \cdot \mathbf{u}] + \nabla \cdot \mathbf{j} = \dot{\omega}_T + S_{energy}, \tag{3}$$

$$\frac{\partial (\rho Y_m)}{\partial t} + \nabla \cdot [\mathbf{u}(\rho Y_m)] + \nabla \cdot \mathbf{s_m} = \dot{\omega}_m + S_{species,m}, (m = 1, \ldots M - 1). \tag{4}$$

In above equations, *t* is time and $\nabla \cdot (\cdot)$ is the divergence operator. $\rho$ is the gas density, and $\mathbf{u}$ is the gas velocity vector. $p$ is the pressure, updated from the equation of state, i.e., $p = \rho RT$. $T$ is the gas temperature. *R* is the specific gas constant., calculated from $R = R_u \sum_{m=1}^{M} Y_m W_m^{-1}$. $W_m$ is the molar weight of *m*-th species and $R_u = 8.314$ J/(mol·K) is the universal gas constant. In Eq. (4), $Y_m$ is the mass fraction of *m*-th species, and $M$ is the total species number. $\boldsymbol{E} \equiv e_s + |\mathbf{u}|^2/2$ is the total non-chemical energy. $e_s = h_s - p/\rho$ is the sensible internal energy and $h_s$ is the sensible enthalpy [33].

The viscous stress tensor $\mathbf{T}$ in Eq. (2) is modelled by $\mathbf{T} = -2\mu[\mathbf{D} - \mathrm{tr}(\mathbf{D})\mathbf{I}/3]$. Here $\mu$ is the dynamic viscosity and follows the Sutherland's law [34]. $\mathbf{D} \equiv [\nabla\mathbf{u} + (\nabla\mathbf{u})^T]/2$ is the deformation gradient tensor. $\mathbf{I}$ denotes the unit tensor, and $\mathrm{tr}(\cdot)$ is the trace of a tensor. In addition, $\mathbf{j}$ in Eq. (3) is the diffusive heat flux modelled with Fourier's law, i.e., $\mathbf{j} = -k\nabla T$. The thermal conductivity $k$ is calculated based on the Eucken approximation [35].

In Eq. (4), $\mathbf{s_m} = -D_m\nabla(\rho Y_m)$ is the species mass flux. The mass diffusivity $D_m$ is calculated through $D_m = k/\rho C_p$ with unity Lewis number assumption, which is deemed reasonable for detonation combustion modelling. $C_p$ is constant pressure heat capacity of the gas phase. Moreover, $\dot{\omega}_m$ is the production or consumption rate of $m$-th species by all $N$ reactions. Also, the term $\dot{\omega}_T$ in Eq. (3) represents the heat release from chemical reactions and is estimated as $\dot{\omega}_T = -\sum_{m=1}^{M} \dot{\omega}_m \Delta h_{f,m}^o$, in which $\Delta h_{f,m}^o$ is the formation enthalpy of $m$-th species.

## 2.2 Liquid phase

The Lagrangian method is used to model the dispersed liquid phase with a large number of spherical droplets [36]. The interactions between droplets are neglected because we only study dilute water sprays with initial droplet volume fraction being generally less than 0.1% [37]. Since the ratio of gas density to the water droplet material density is well below one, the Basset force, history force, and gravity force are not considered [37]. Therefore, the governing equations of mass, momentum, and energy for a single water droplet read

$$\frac{dm_d}{dt} = -\dot{m}_d, \tag{5}$$

$$\frac{d\mathbf{u}_d}{dt} = \frac{\mathbf{F}_d + \mathbf{F}_p}{m_d}, \tag{6}$$

$$c_{p,d}\frac{dT_d}{dt} = \frac{\dot{Q}_c + \dot{Q}_{lat}}{m_d}, \tag{7}$$

where $m_d = \pi\rho_d d_d^3/6$ is the mass of a single droplet, and $\rho_d$ and $d_d$ are the droplet material density and diameter, respectively. $\mathbf{u}_d$ is the droplet velocity vector, $\mathbf{F}_d$ and $\mathbf{F}_p$ are the drag and pressure gradient force acting on the droplet. $c_{p,d}$ is the droplet heat capacity at constant pressure, and $T_d$ is the droplet temperature. In this work, both $\rho_d$ and $c_{p,d}$ are dependent on the droplet temperature $T_d$ [38].

The evaporation rate, $\dot{m}_d$, in Eq. (5) is

$$\dot{m}_d = k_c A_d W_d (c_s - c_g), \tag{8}$$

where $A_d$ is the surface area of a single droplet, $k_c$ is the mass transfer coefficient, and $W_d$ is the molecular weight of the vapor. $c_S$ is the vapor mass concentration at the droplet surface, i.e.

$$c_S = \frac{p_{sat}}{R_u T_f}, \tag{9}$$

where $p_{sat}$ is the saturation pressure. The droplet surface temperature $T_f$ is calculated from $T_f = (T + 2T_d)/3$ [39]. In Eq. (8), the vapor mass concentration in the gas phase, $c_g$, is

$$c_g = \frac{p x_i}{R_u T_f}, \tag{10}$$

where $x_i$ is the water vapor mole fraction in the bulk gas.

The mass transfer coefficient, $k_c$, in Eq. (8) is calculated from [40]

$$Sh = \frac{k_c d_d}{D_f} = 2.0 + 0.6 Re_d^{1/2} Sc^{1/3}, \tag{11}$$

where $Sh$ is the Sherwood number, $D_f$ is the vapor mass diffusivity in the gas phase [41], and $Sc$ is the Schmidt number of gas phase. The droplet Reynolds number, $Re_d$, is defined as

$$Re_d \equiv \frac{\rho d_d |\mathbf{u}_d - \mathbf{u}|}{\mu}. \tag{12}$$

The drag force $\mathbf{F}_d$ in Eq. (6) is modelled as [42]

$$\mathbf{F}_d = \frac{18\mu}{\rho_d d_d^2} \frac{C_d Re_d}{24} m_d (\mathbf{u} - \mathbf{u}_d). \tag{13}$$

The drag coefficient, $C_d$, is estimated with [42]

$$C_d = \begin{cases} 0.424, & if\ Re_d \geq 1000, \\ \frac{24}{Re_d}\left(1 + \frac{1}{6} Re_d^{2/3}\right), & if\ Re_d < 1000. \end{cases} \tag{14}$$

Besides, the pressure gradient force $\mathbf{F}_p$ in Eq. (6) takes the following form

$$\mathbf{F}_p = -V_d \nabla p. \tag{15}$$

Here $V_d$ is the volume of a single water droplet.

The convective heat transfer rate $\dot{Q}_c$ in Eq. (7) is

$$\dot{Q}_c = h_c A_d (T - T_d). \quad (16)$$

Here $h_c$ is the convective heat transfer coefficient, following Ranz and Marshall [40]

$$Nu = h_c \frac{d_d}{k} = 2.0 + 0.6 Re_d^{1/2} Pr^{1/3}, \quad (17)$$

where $Nu$ and $Pr$ are the Nusselt and Prandtl numbers of gas phase, respectively. In addition, $\dot{Q}_{lat}$ in Eq. (7) is the latent heat absorption by droplet evaporation.

The droplet effects on the gas phase are modelled with Particle-source-in-cell approach [43], which are realized with the source terms, $S_{mass}$, $\mathbf{S}_{mom}$, $S_{energy}$ and $S_{species,m}$, for the gas phase equations (1)−(4). They are calculated from

$$S_{mass} = \frac{1}{V_c} \sum_{i=1}^{N_d} \dot{m}_{d,i}, \quad (18)$$

$$\mathbf{S}_{mom} = -\frac{1}{V_c} \sum_{i=1}^{N_d} (-\dot{m}_{d,i} \mathbf{u}_{d,i} + \mathbf{F}_{d,i} + \mathbf{F}_{p,i}), \quad (19)$$

$$S_{energy} = -\frac{1}{V_c} \sum_{i=1}^{N_d} \left(-\dot{m}_{d,i} h_v + \dot{Q}_{c,i}\right), \quad (20)$$

$$S_{species,m} = \begin{cases} S_{mass} & for\ H_2O\ species \\ 0 & for\ other\ species. \end{cases} \quad (21)$$

$V_c$ is the CFD cell volume and $N_d$ is the droplet number in one CFD cell. The term $-\dot{m}_d \mathbf{u}_d$ is the momentum transfer due to droplet evaporation, whereas $h_v$ is the water vapor enthalpy at the droplet temperature $T_d$. Hydrodynamic force work and kinetic energy carried by the vapor are not considered in Eq. (20), because they are 2-3 orders of magnitude smaller compared to other energy exchange terms for spray detonations [44].

### 2.3 Computational method

For the gas phase equations, second-order backward method is employed for temporal discretization and the time step is about $1\times10^{-11}$ s. The MUSCL-type Riemann-solver-free scheme by Kurganov et al. [45] with van Leer limiter is used for convective flux calculations in momentum equations. Total variation diminishing scheme is used for the convection terms in energy and species equations. Also, second-order central differencing scheme is applied for the diffusion terms in Eqs. (2)−(4). The mechanism (13 species and 27 reactions) by Burke et al. [46] is used for predicting hydrogen detonative combustion.

For the liquid phase, the water droplets are tracked based on their barycentric coordinates. The equations, i.e., Eqs. (5)−(7), are integrated by first-order implicit Euler method. Meanwhile, the gas properties (e.g., velocity and temperature) at the droplet location are calculated based on linear interpolation.

The above numerical methods and the sub-models depicted in Section 2.2 have been carefully validated and verified for detonation problems in gaseous and gas−droplet two-phase flows [30,31]. They have been successfully applied for various detonation and supersonic combustion problems [25,31,47–52].

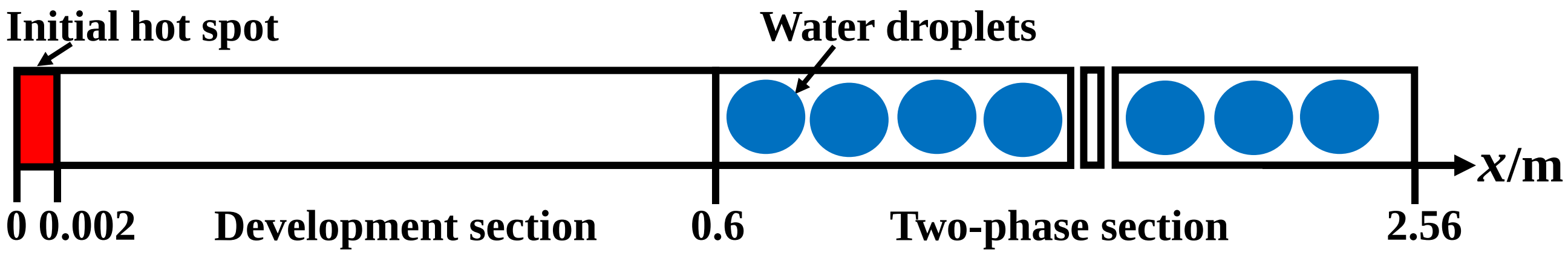

Figure 1 Schematic of the 1D computational domain. Domain and droplet sizes not to scale.

## 3. Physical model and numerical implementation

Incident hydrogen detonation in one-dimensional (1D) domain laden with fine water droplets are studied in this work, see Fig. 1. The 1D domain is 2.56 m long ($x$-direction), including detonation development and gas−liquid two-phase sections. The domain is initially filled with argon diluted stoichiometric $H_2/O_2$ premixture (molar ratio of $H_2/O_2/Ar$: 2/1/7). The initial gas temperature and pressure are $T_0$ = 300 K and $p_0$ = 0.2 atm, respectively. Uniform Cartesian cells of 20 μm are used for the whole domain, and the resultant total cell number is 128,000. The Half-Reaction Length (HRL) estimated from the ZND structure of droplet-free $H_2/O_2/Ar$ detonation is $\Delta_{HRL}$ ≈ 478 μm. Considering the fact that dispersed water droplets may thicken the induction zone length due to heat and/or momentum transfer [9,44], there are at least 23 cells for $\Delta_{HRL}$ in our spray detonation simulations. Mesh resolution analysis is provided in Section A of Supplementary Document and the results confirm the sufficiency of mesh size for the gas phase.

The detonation wave is initiated by a hot spot (2,000 K and 20 atm) near the left end (see Fig. 1). For the left boundary ($x$ = 0), the non-reflective condition is enforced for the pressure, while the zero gradient condition for other quantities [53]. Moreover, zero gradient conditions are assumed for the right boundary at $x$ = 2.56 m.

Monodispersed and static ($\mathbf{u}_d = 0$) water droplets are uniformly distributed in the two-phase section (i.e., $x$ = 0.6 — 2.56 m). The initial water mass loading of $z$ = 0.01 — 0.5 and droplet sizes of $d_d^0$ = 2 — 15 μm will be studied in this work. Note that the mass loading is defined based on the gas properties in the two-phase section (i.e., $x$ > 0.6 m). The initial temperature, material density and isobaric heat capacity of the water droplets are 300 K, 997 kg/m$^3$ and 4,187 J/kg·K, respectively.

The computational parcel is used in the simulations, which is an ensemble of the water droplets with the same properties (e.g., velocity, size, and temperature). Initially each cell has one parcel in the simulations and accordingly the water droplet number in each parcel, $n_p$, is varied based on different mass loadings. This method is sufficient for capturing droplet evolutions and interphase coupling, based on our sensitivity analysis in Section B of Supplementary Document.

In light of the foregoing Eulerian mesh size $\Delta$ (20 μm) and initial droplet diameters $d_d^0$ (2 — 15 μm), their ratios, $\varepsilon \equiv \Delta/d_d^0$, range from 1.33 to 10. From the tests of single droplet evaporation in stationary gas [54,55], it is shown that $\varepsilon \geq 10$ is required to have accurate predictions of the droplet evaporation. Recently, Kronenburg and his co-workers [56] make relative error estimations of droplet evaporation timescale $\epsilon_\tau$, considering more physical and numerical factors, such as cell and droplet size ratio $\varepsilon$, initial liquid mass loading $z'$, cell Pélect number $Pe_{\Delta x}$, and Sherwood number $Sh'$. They derive the following estimations of evaporation timescale error $\epsilon_\tau$ [56]:

$$\epsilon_\tau \approx min\left[\varepsilon^{-1}, z', \frac{2\pi}{3}\varepsilon^{-1}\left(\frac{Pe_{\Delta x}}{Sh'}\right)^{-1}\right]. \tag{22}$$

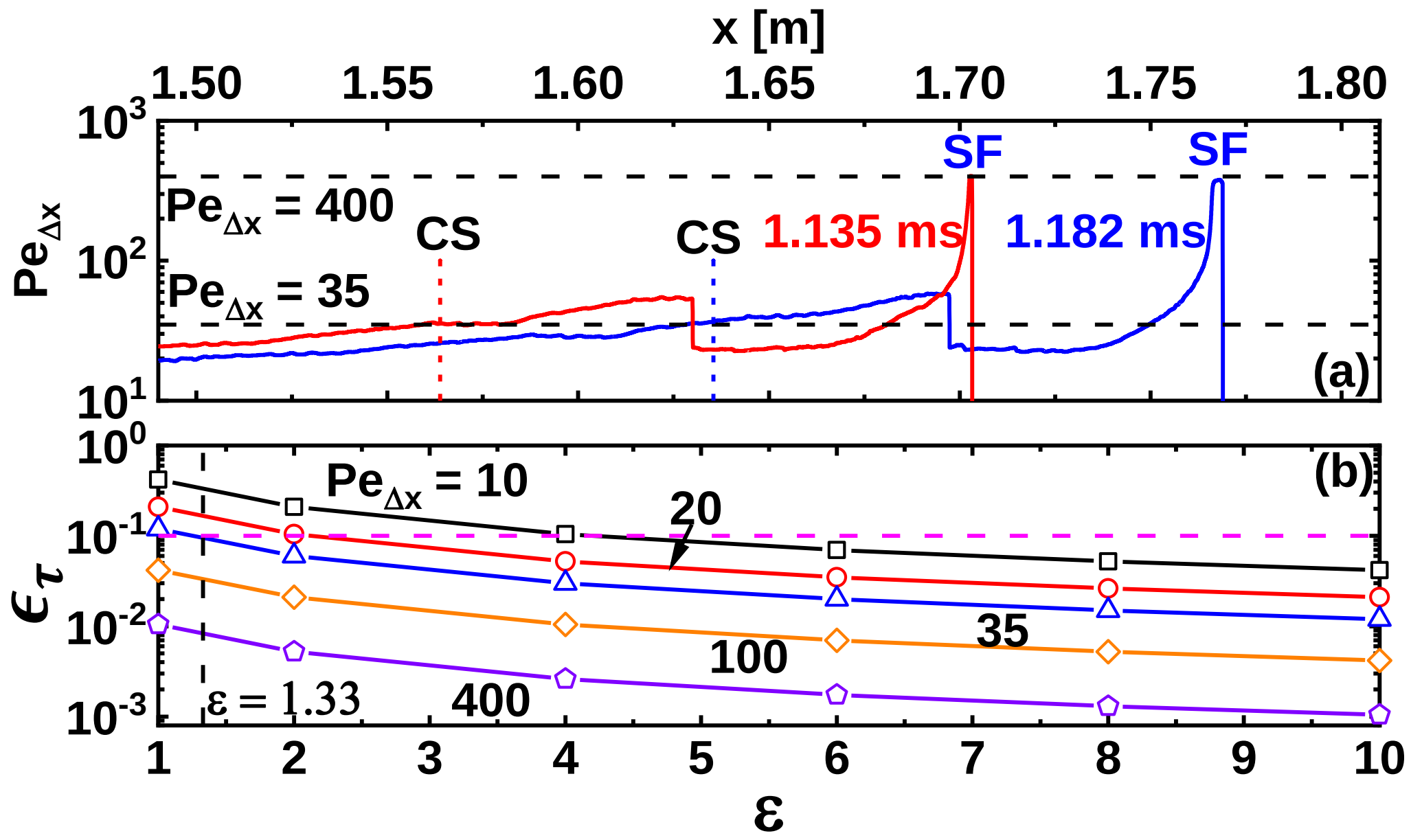

Figure 2 (a) Spatial evolution of cell Pélect number $Pe_{\Delta x}$, and (b) evaporation timescale error $\epsilon_\tau$ as a function of cell / droplet size ratio $\varepsilon$. CS: two-phase contact surface. SF: shock front. Results from case: $d_d^0$ = 5 μm and $z$ = 0.07.

To estimate $\epsilon_\tau$ following Eq. (22), we calculate the cell Pélect number $Pe_{\Delta x}$ at two instants from one of our simulated cases ($d_d^0$ = 5 μm and $z$ = 0.07) in which detonation can stably propagate and the results are in Fig. 2. It is observed that cell Pélect number $Pe_{\Delta x}$ roughly varies from 35 to 400 within the evaporating droplet areas behind the detonation wave (between the SF and two-phase Contact Surface (CS), marked in Fig. 2a). Note that here the two-phase CS is the interface between the gas-only and gas-liquid mixtures. Based on the results in Fig. 2(a), $\frac{2\pi}{3}\varepsilon^{-1}\left(\frac{Pe_{\Delta x}}{Sh'}\right)^{-1} < \varepsilon^{-1}$ is generally valid. Figure 2(b) shows the evaporation timescale error $\epsilon_\tau$ from the tighter conditions in Eq. (22), i.e., the last expression, considering the cell Pélect numbers of 10−400. For simplicity, $Sh' = 2$ is assumed. Apparently, inclusion of the cell Pélect number considerably reduces the requirement of $\varepsilon$ to achieve an equivalent $\epsilon_\tau$. For example, for $Pe_{\Delta x}$ = 35, to achieve $\epsilon_\tau$ < 10%, $\varepsilon$ > 1.33 (marked in Fig. 2b) is required, which is much more relaxed compared to the requirement of $\varepsilon$ < 10 derived from single droplet evaporation in stationary atmosphere ($Pe_{\Delta x}$ = 0), which is essentially a diffusion-controlled process [54,55]. This confirms the important influences of strong convection flows (such as detonations) in modelling droplet evaporation rate. Physically, it means that the vapour released from the droplets can be swept away quickly by local convection, which considerably affects the vapour concentration difference near the droplet surface. In general, based on the above analysis, the evaporation timescale errors in our simulations (1.33 < $\varepsilon$ < 10) are largely less than 10%. This is also favorable for the applicability of the Eulerian−Lagrangian approach in highly resolved detonation simulations with dispersed droplets.

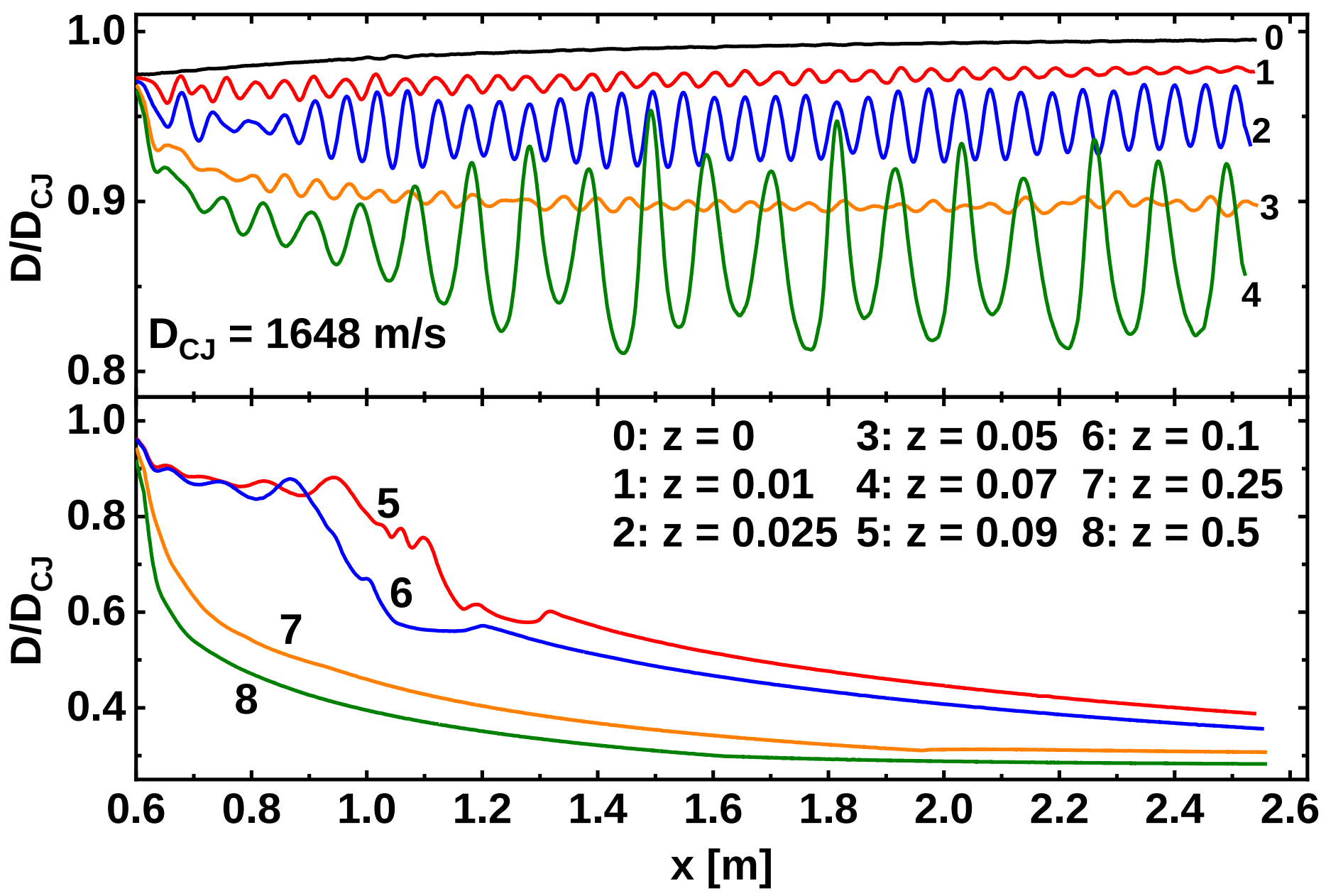

Figure 3 Evolutions of leading shock propagation speed with various water mass loadings. $d_d^0$ = 5 μm.

## 4. Results and discussion

### 4.1 Detonation propagation mode in water cloud

Figure 3 shows the propagation speed of the leading shock front with various water mass loadings. The initial droplet diameter is fixed to be $d_d^0$ = 5 μm. The results are scaled by the C−J speed, $D_{CJ}$ = 1648 m/s, of the droplet-free case ($z$ = 0, or interchangeably termed as dry mixture). One can see that, after the detonation wave enters the fine water sprays ($x$ > 0.6 m), three unsteady propagation modes can be identified: (1) pulsating propagation ($z$ = 0.01-0.07), (2) extinction after a finite propagation ($z$ = 0.09 and 0.1), and (3) immediate extinction ($z$ = 0.25 and 0.5).

In the first mode (lines #1−#4 in Fig. 3), oscillating shock wave speeds are observed in the two-phase section. For instance, when $z$ = 0.025, the speed varies between 1,520 and 1,585 m/s, whilst the pulsation frequency is approximately 28,409 Hz. The speeds in the two-phase cases are lower than that of the water-free case (line #0). In the latter, the propagation speed increases monotonically as the detonation wave propagates and is very close to the C—J speed ($D/D_{CJ} \rightarrow 1$). This indicates the significant effects of evaporating water droplets on the incident detonations. It is also seen that the magnitudes and frequencies are not monotonic with water mass loading. When $z$ = 0.01 and 0.05 (lines #1 and #3), the oscillation magnitudes are much lower, compared to those with $z$ = 0.025 and 0.07. This will be further discussed within a wider parameter range in Section 4.4. Moreover, it is observed in Fig. 3 that propagation speed decreases with increased water mass loading. For example, the mean detonation propagation speeds are 1,554 and 1,434 m/s for $z$ = 0.025 and 0.07, respectively. This may be because larger water mass loading has stronger weakening effect as it can absorb more energy and momentum from gas phase.

In the second mode with higher mass loading (e.g., 0.09 and 0.1, lines #5 and #6), non-monotonic changes of the shock propagation speed are present when the detonation just encroaches the two-phase section, but there are no pulsating detonations followed. After travelling a finitely long distance (about $x$ = 1 m), the shock propagation speed decays quickly, indicating the reduced interactions with the chemical reactions due to decoupling of the leading Shock Front (SF) and Reaction Front (RF). For the third mode ($z$ = 0.25 and 0.5, lines #7 and #8), decoupling of the SF and RF occurs earlier and the detonation is quenched immediately after the detonation wave arrives at the heavier water cloud, which is manifested by the monotonically decreasing shock propagation speed in the water spray area.

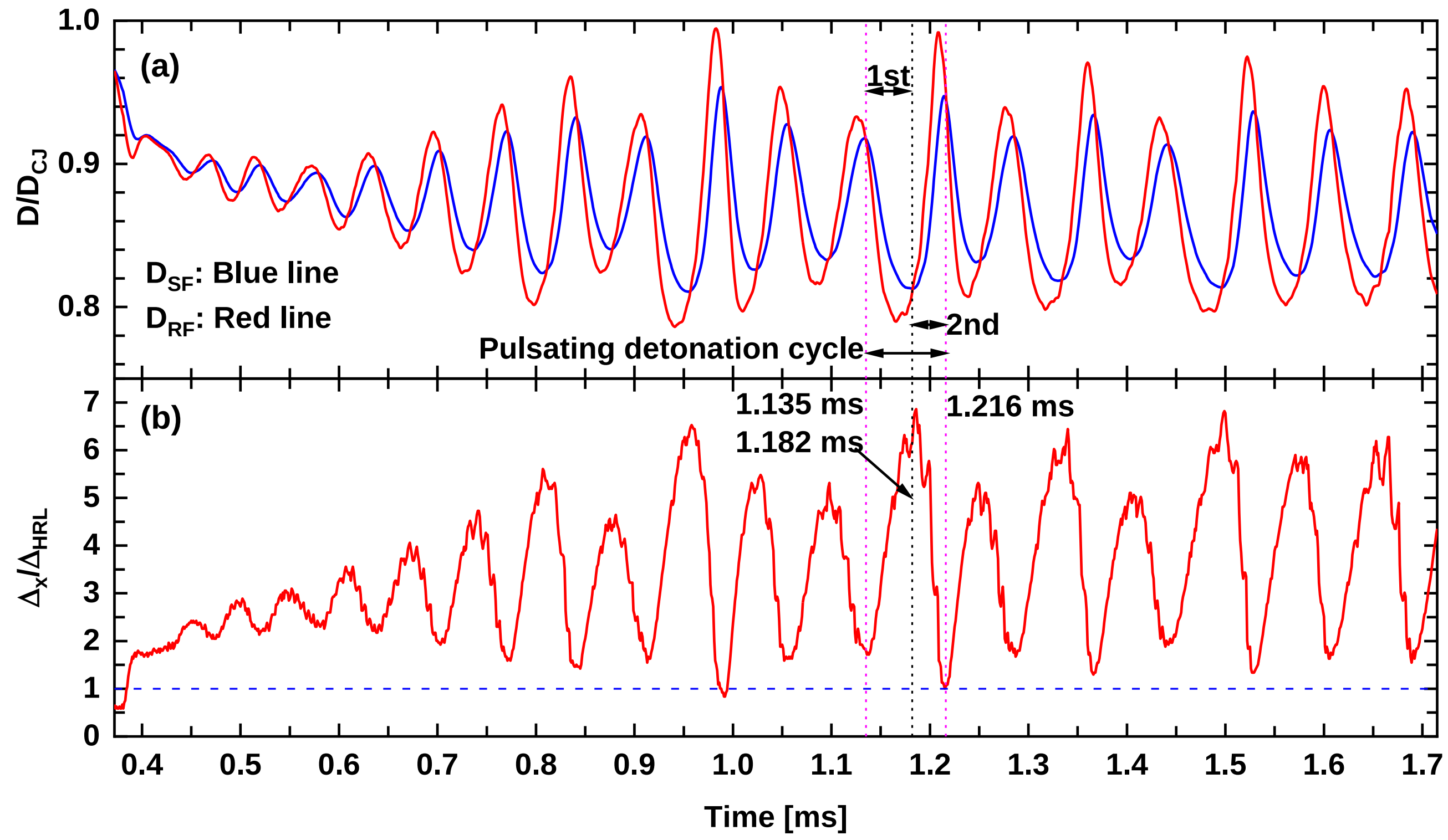

Figure 4 Time history of (a) propagation speed of shock front $D_{SF}$ and the reaction front $D_{RF}$, as well as (b) the distance $\Delta_x$ between the two fronts. $d_d^0$ = 5 μm and $z$ = 0.07.

## 4.2 Pulsating detonation

To articulate the pulsating detonation dynamics in the first mode, the propagation speeds of RF ($D_{RF}$) and SF ($D_{SF}$) are plotted in Fig. 4(a), which corresponds to $z$ = 0.07 in Fig. 3. Note in passing that the detonation wave enters the two-phase section at 0.371 ms. After an initial transition, both RF and SF propagation speeds exhibit periodic fluctuations, but with a small phase difference. This becomes more regular since 0.7 ms. The propagation speeds of both fronts in water sprays are below than the C-J speed of the dry mixture. The pulsation behavior is accompanied by the periodic variation of the distance between the RF and SF, $\Delta_x$, as shown in Fig. 4(b). In general, $\Delta_x$ is higher than $\Delta_{HRL}$ from the droplet-free case, which is caused by the heat absorption and/or momentum exchange of the evaporating droplets.

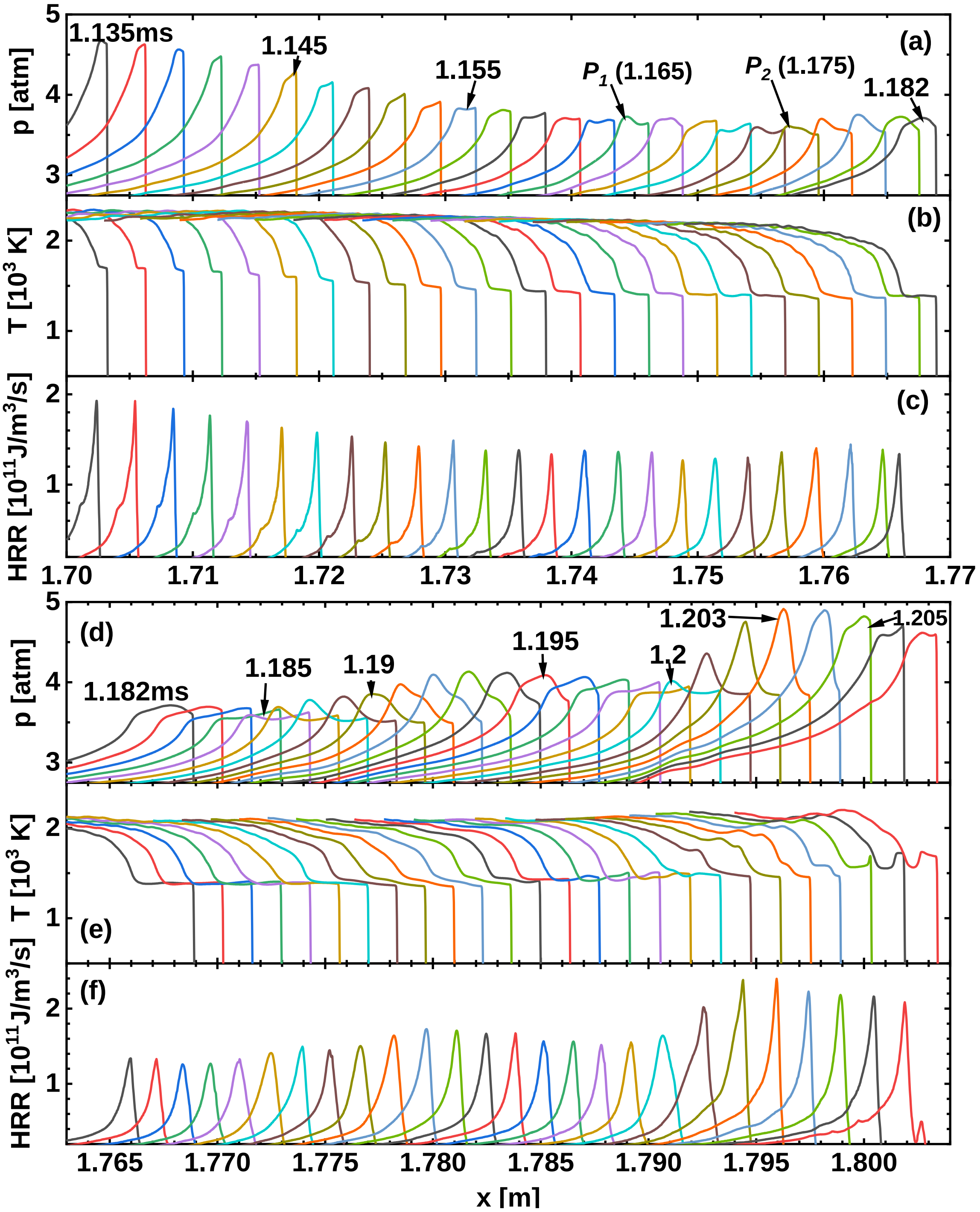

Figure 5 Profiles of (a, d) pressure, (b, e) temperature, and (c, f) heat release rate within one pulsating detonation cycle marked in Fig. 4. The 1st stage: (a)-(c); 2nd stage: (d)-(f). $d_d^0$ = 5 μm and $z$ = 0.07.

One pulsation cycle, e.g., 1.135 ms < $t$ < 1.216 ms, can be divided into first and second stages, based on the maxima and minima of the front distance $\Delta_x$. At the beginning of the first cycle (1.135 ms < $t$ < 1.182 ms), both $D_{RF}$ and $D_{SF}$ are high, with relatively small front distance of $\Delta_x \approx 2.0\Delta_{HRL}$. Then SF and RF decelerate, but the SF speed is still larger than the RF one. Accordingly, their distance $\Delta_x$ continuously increases to roughly seven times $\Delta_{HRL}$. This indicates that the RF and SF are experiencing decoupling in this period (see 1st stage in Fig. 4). Moreover, in the second half stage (i.e., 1.182 ms < $t$ < 1.216 ms), the speeds of both RF and SF increase, and the RF speed $D_{SF}$ is always higher than the SF one $D_{RF}$. The RF gradually catches up with the leading shock and this corresponds to RF and SF coupling. At 1.216 ms, their distance is close to $\Delta_{HRL}$. Nonetheless, such coupling is almost instantaneous, and the RF and SF start the next pulsation cycle with gradually increased $\Delta_x$, which leads to a galloping movement when the detonation wave propagates in the water spray cloud.

Figure 5 shows the evolutions of the gas pressure, temperature, and Heat Release Rate (HRR) within the pulsation cycle discussed above. For the first stage of the cycle in Figs. 5(a)−5(c), the decoupling of the SF and RF is characterized by gradual reduction of shock pressure, gas temperature, and maximum HRR. A key phenomenon observed from Figs. 5(a) is that multiple pressure waves (or reaction shock [12]), e.g., labelled with $P_1$ and $P_2$, are generated due to the strong heat release from the RF. These new pressure waves can surge forward in the induction zone and ultimately overtake the lead SF, thereby continuously enhancing the leading shock intensity, as seen from the last several profiles of the 1st stage from Fig. 5(a). The wave interactions between the RF and SF are also reported in previous studies on gaseous galloping detonations [12,15,17,18,25]. In the 2nd stage in Figs. 5(c)−5(d), this phenomenon repeatedly occurs, and the leading shock pressure increases quickly from 1.182 ms to 1.203 ms. The mutual enhancement between the RF and pressure wave becomes strong after 1.2 ms, featured by increased peak pressure and HRR. One can see that in one cycle multiple pressure waves are generated. This may be because the thermodynamic conditions in the induction zone is affected by heat absorption and lead shock attenuation by the droplets. Unlike the pulsation mechanism discussed by Yungster and Radhakrishnan for $H_2$/air detonations [15], single pressure wave is not sufficient to enhance the shocked gas reactivity above the autoignition threshold to induce localized hot spot and deflagration-to-detonation transition.

The detailed evolutions between 1.203 and 1.209 ms are presented in Fig. 6. A pressure wave emanated from the RF is clearly observed at 1.203 ms, which penetrates the leading SF at 1.206 ms and the shock pressure is hence increased by 22.4% compared to 1.203 ms. Meanwhile, a rarefaction wave and downstream facing contact surface are generated, and the latter demarcates the relatively hot and cold shocked gas. At 1.207 ms, one autoigniting spot (marked as AS) arises near the contact surface and at the foot of the RF, which corresponds to increased HRR, temperature, and consumption of the local hydrogen. This location is more favourable for autoignition, because of higher temperature behind the contact surface [57] and radical diffusion from the RF. Double HRR peaks appear behind the leading SF: the left peak is from the original detonative combustion, whilst the right is induced by combustion of the shocked $H_2/O_2/Ar$ mixture.

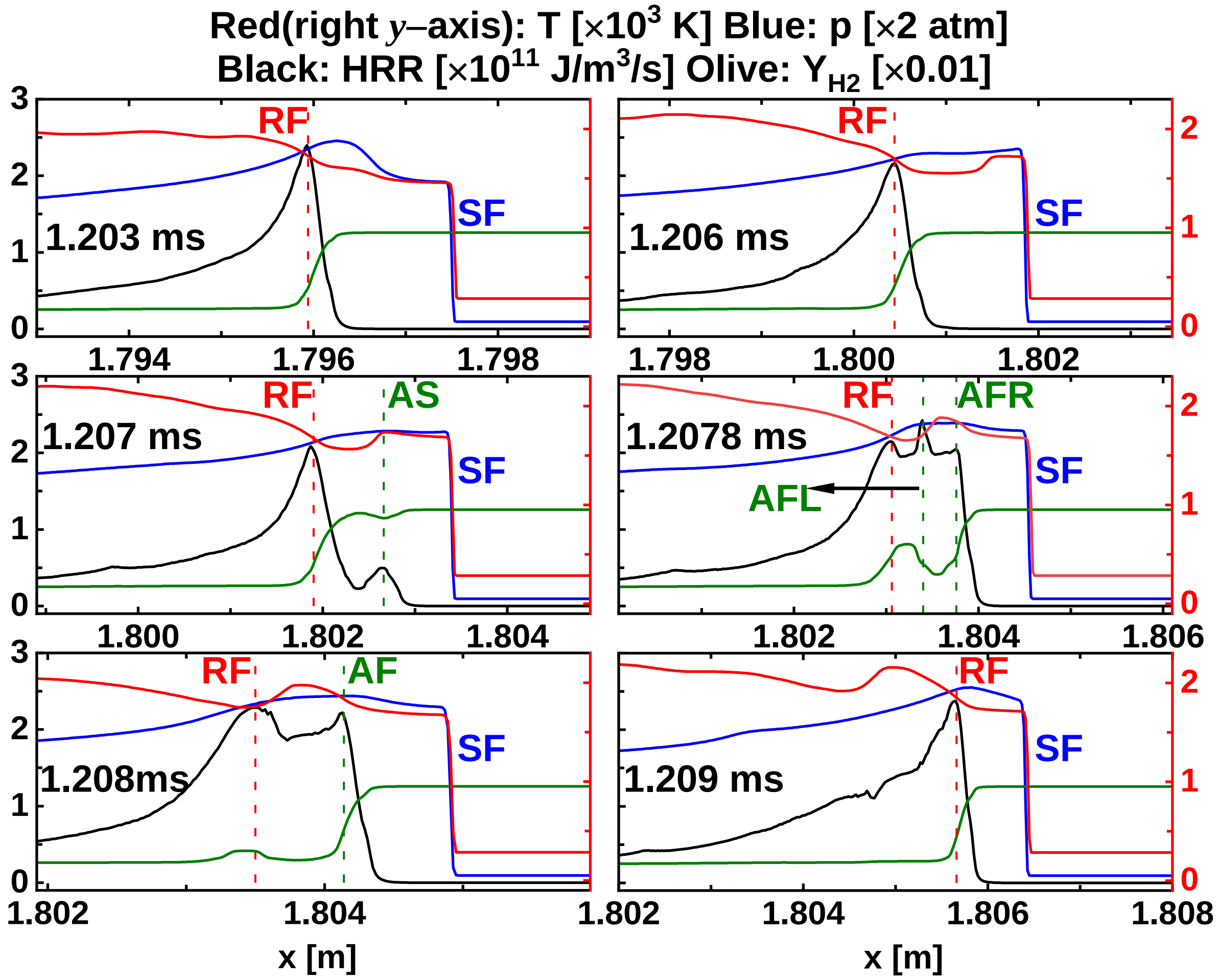

Figure 6 Profiles of pressure, temperature, heat release rate, and $H_2$ mass fraction within one pulsating cycle. Red and olive dashed lines: RF and AS/AF (Autoignition Spot/Front). $d_d^0$ = 5 μm and $z$ = 0.07.

At 1.2078 ms, left- and right-running (in the shock frame) autoigniting fronts (AFL/AFR) evolve from the AS, resulting in three heat release peaks. The AFL consumes the mixture ahead of RF and collides with the RF, leading to reduced heat release. The original RF ultimately fades, and the AFR induced by the reactive spot gradually becomes strong and propagates towards the leading SF at 1.209 ms. Therefore, this leads to a relay between the two RFs in one pulsation cycle. The distance between the new RF and SF approaches $\Delta_{HRL}$ at the end of the pulsating detonation cycle, as shown in Fig. 4(b). However, at this instant, the RF speed is lower than that of the SF, and therefore their distance would increase again, to initiate an ensuing cycle.

The evolutions of the thermochemical structure in Fig. 6 are further quantified through chemical explosive mode analysis [26–28,58,59]. Detailed information about this method can be found in Refs. [26,29,59,60]. Figure 7 shows the spatial evolutions of eigen values of the chemical Jacobian and explosion indices (EI) for temperature and key radicals between the RF and SF at four instants. The EI is calculated based on the eigenvectors associated with the chemical explosive mode [60–62], i.e., $\mathbf{EI} = |\mathbf{a_e} \otimes \mathbf{b_e^T}| / \sum |\mathbf{a_e} \otimes \mathbf{b_e^T}|$, where "⊗" denotes element-wise multiplication of two vectors. Larger radical EI suggests the higher significance of the chemical runaway (chain-branching reactions) in local chemical reaction, whilst thermal runaway proceeds if the gas temperature plays a more important role [26]. Gas temperature profiles are also presented for reference, which are colored by logarithmic expression of the eigen value of chemical Jacobian $\lambda_e$, i.e., $\lambda_{CEM} \equiv sign[Re(\lambda_e)] \cdot log_{10}[1 + |Re(\lambda_e)|]$. Positive (negative) $\lambda_e$ indicates chemical explosive mixtures (non-reactive or burned mixtures). At 1.207 ms in Fig. 7(a), the entire induction zone is chemically explosive, with high value of $\lambda_{CEM}$. Most of it is controlled by the chain-branching reactions with radical H having the highest EI, featuring a chemical runaway process. Thermal runaway is only observed ahead of the RF. However, near the AS location, the importance of the gas temperature becomes greater, although it is still lower than some radical EIs, such as OH and O. At 1.2076 ms, temperature EI becomes increasingly important near the AS, indicating the developing thermal runaway area. At 1.2078 ms, the burned state is presented behind the two autoigniting fronts, i.e., AFL and AFR, as shown in the temperature profile with $\lambda_{CEM} < 0$ in Fig. 7. Accordingly, the induction zone is divided into two un-connected chemically explosive areas, i.e., behind SF and ahead of the RF (see warm color of the gas temperature curve in Fig. 7c). The downstream one gradually shrinks as the fuel is consumed by the RF and AFL. At 1.208 ms, only the chemical explosive area between the AF and SF is observed, which is the new induction zone between the SF and new RF.

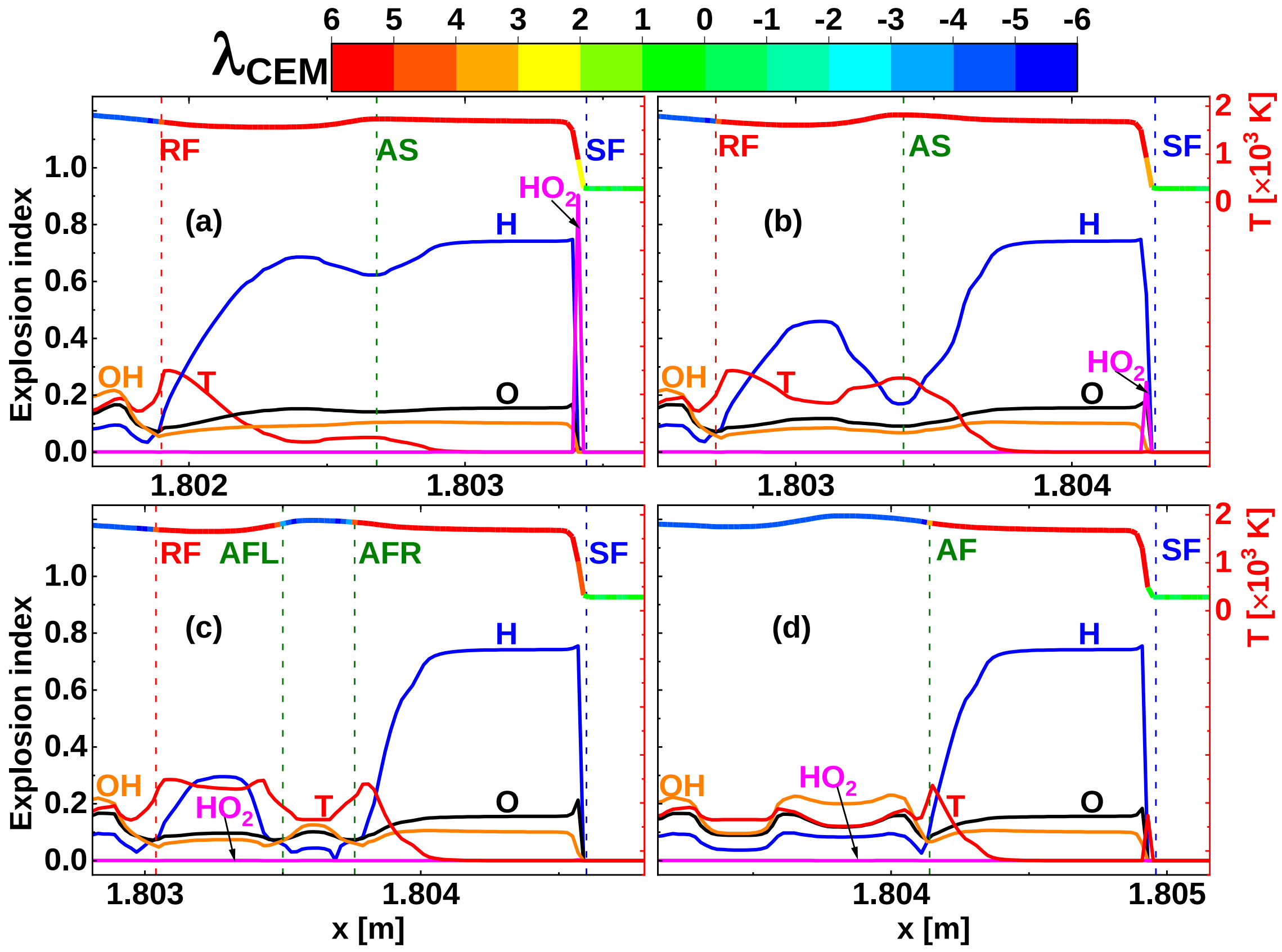

Figure 7 Evolutions of explosion indices for species and temperature between the RF and SF at four instants: (a) 1.207, (b) 1.2076, (c) 1.2078, and (d) 1.208 ms. Gas temperature curves are colored by $\lambda_{CEM}$. $d_d^0$ = 5 μm and $z$ = 0.07.

The flow field structure of pulsating detonation in water sprays is analyzed in Fig. 8, through the spatial profiles of gas temperature, velocity, shock-frame Mach number, and droplet diameter at 1.182 ms. The shock-frame subsonic zone ends at the sonic point (SP), and the gas−liquid two-phase zone ends at about $x$ = 1.635 m (marked at CS). It can be observed that due to lead shock compression, the gas temperature experiences the first increase at $x$ = 1.769 m, followed by a thermally neutral zone behind SF. Then chemical reaction starts at RF and the gas temperature is increased to more than 2,000 K. The water droplets are heated to saturation temperature within a finite distance (about 0.006 m) behind the SF (marked as STP). The temperature equilibrium between the gas phase and water droplets cannot be achieved; instead, large temperature difference exists and hence strong convective heat transfer proceeds in the detonated area. Moreover, the static droplets respond quickly to the high-speed flows because of short droplet momentum relaxation time, and their velocity increases behind the leading SF. Velocity quasi-equilibrium is reached (~ 600 m/s) at about $x$ = 1.745 m (marked as VqEP in Fig. 8). Further downstream, relatively small slip velocity always exists, which is caused by the tempo-spatial evolutions of the gas velocities behind the detonation wave and finite droplet momentum relaxation time.

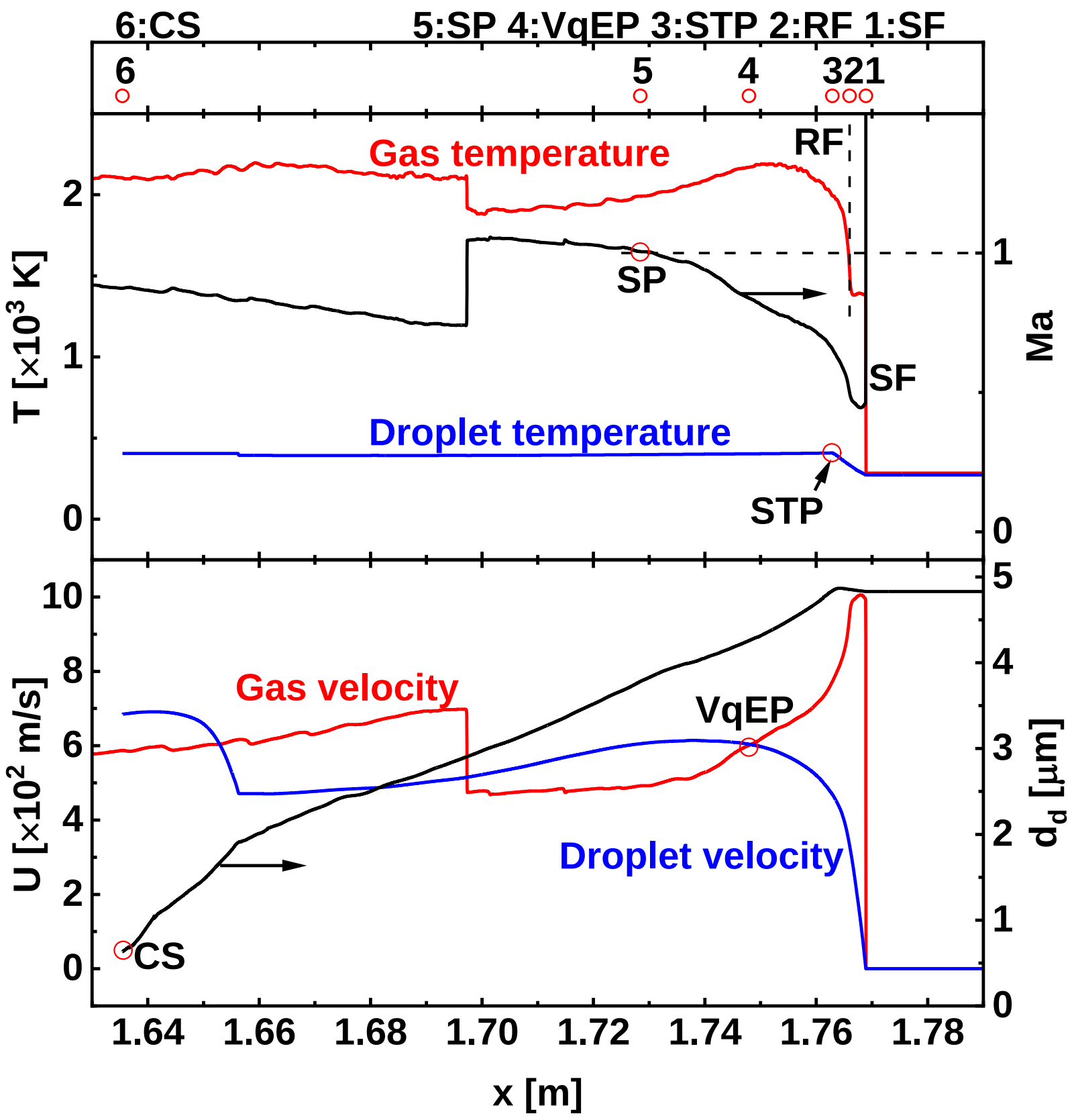

Figure 8 Distribution of temperature, shock-frame Mach number, velocity, and droplet diameter from the case of $d_d^0$ = 5 μm and $z$ = 0.07 at 1.182 ms. SP: sonic point; STP: saturation temperature point; VqEP: velocity quasi-equilibrium point.

The characteristic locations in spray detonations at 1.182 ms, including SF, RF, sonic point, and two-phase contact surface, are presented collectively at the top of Fig. 8. How these locations evolve in a pulsation detonation is shown in Fig. 9. Before (after) the middle dashed line, which corresponds to 1.182 ms, is the first (second) stage of the cycle. It is found that their relative positions are generally unchanged within a single pulsating period, but their distances have pronounced variations. Specifically, the induction zone length, i.e., $\Delta_x = x_{SF} - x_{RF}$, first increases and then decreases, varying between $2.0\Delta_{HRL}$ and $6.0\Delta_{HRL}$, which is already shown in Fig. 4. The similar trend is also observed in the distance of droplet saturation temperature point ($x_{SF} - x_{STP}$) relative to the shock front. This is reasonable because the gas temperature behind the SF is reduced (increased) as RF gradually decouples from (couples with) the SF in the first (second) stage.

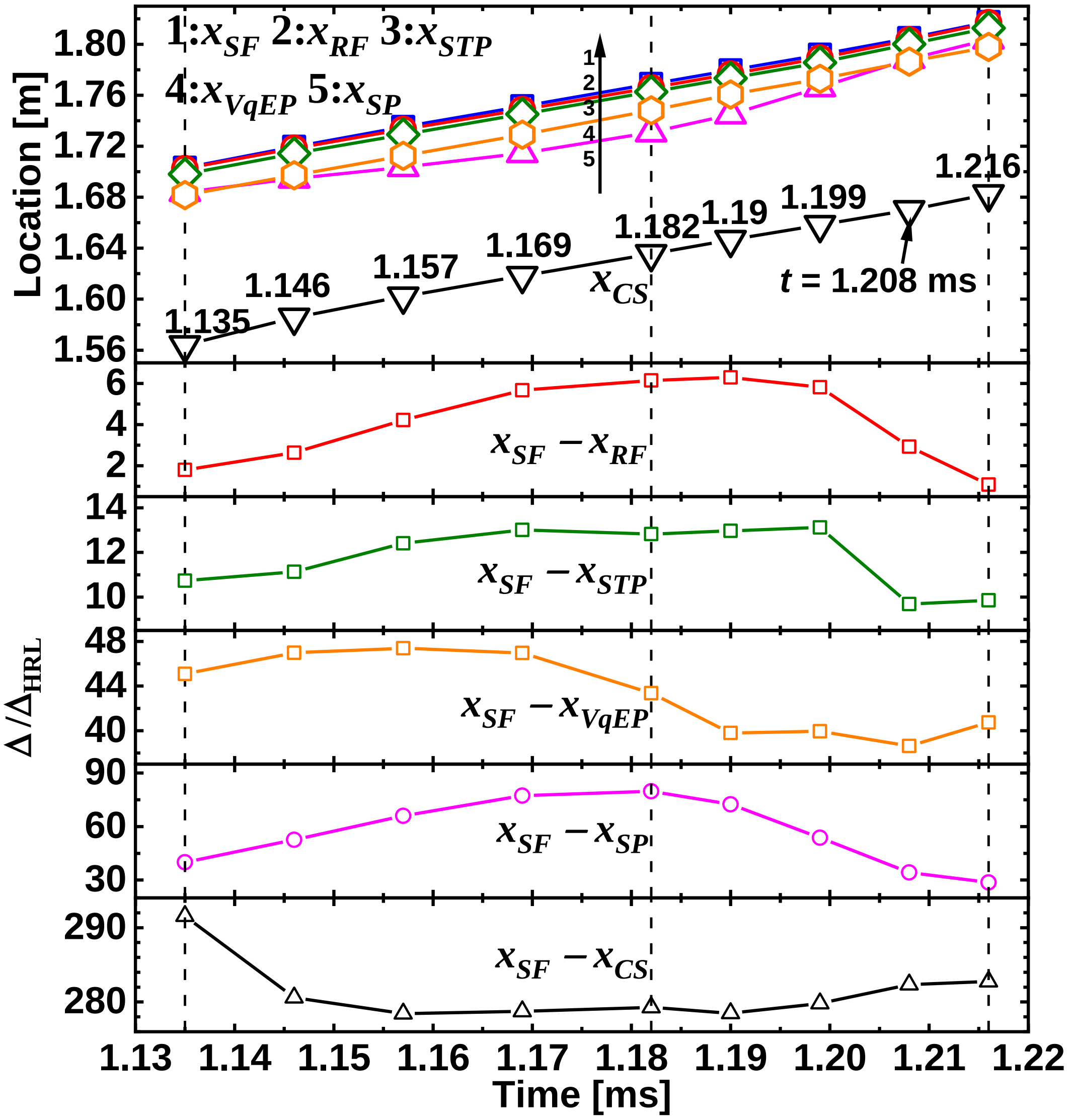

Figure 9 Spatial evolutions of SF, RF, shock-frame sonic point, two-phase contact surface, and their distances within one pulsating cycle. $d_d^0$ = 5 μm and $z$ = 0.07.

However, the velocity quasi-equilibrium point behaves differently, with obvious delay relative to the induction zone length variation. The maximum and minimum values of $x_{SF} - x_{VqEP}$ occur inside the first and second stages, respectively. This is associated with the longer velocity response time of the water droplets. Besides, the length ($x_{SF} - x_{SP}$) of the shock-frame subsonic zone also follows the evolution trend of the induction zone length, but with larger length scales ($40{-}80\Delta_{HRL}$). All the above characteristic locations are in the droplet-laden area. The droplet heating / evaporation zone length in the shocked gas, $x_{SF} - x_{CS}$, also changes (between 275 and $295\Delta_{HRL}$) in the pulsation cycle. It decreases when the pulsation cycle starts, levels off for most of the period, followed by a slow increase at the end of the pulsation cycle. The increase around the end (or the beginning) of the cycle is caused by the accelerating leading shock front and therefore more water droplets are dispersed behind the SF.

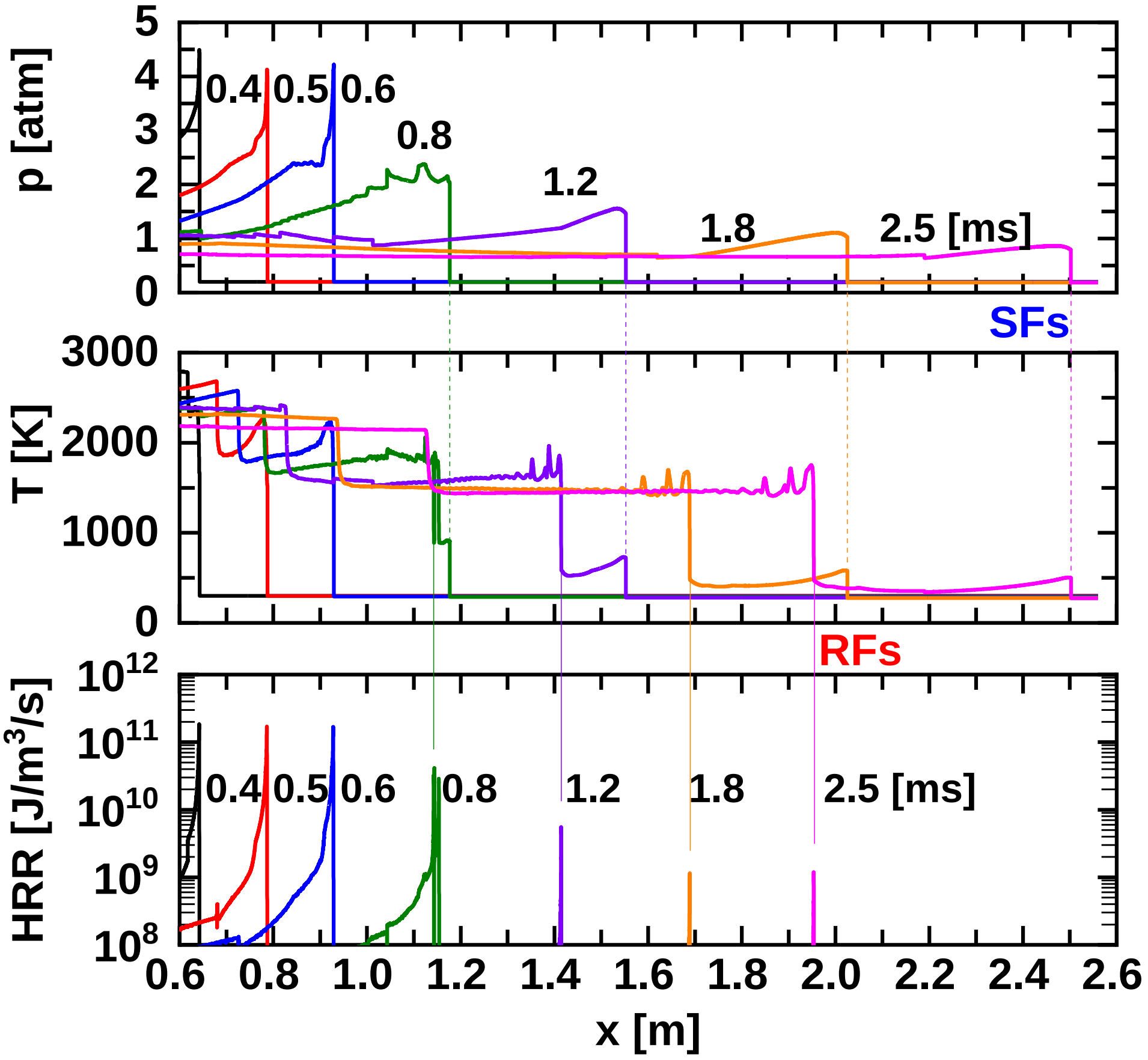

Figure 10 Spatial evolution of pressure, gas temperature, and heat release rate. Dashed lines: SF or RF. $d_d^0$ = 5 μm and $z$ = 0.09.

### 4.3 Detonation extinction

To explore the unsteady physics behind detonation failure in the second and third modes, two cases with mass loadings $z$ = 0.09 and 0.25 (lines #5 and #7 in Fig. 4) are selected for discussion.

Figures 10 and 11 present their time sequence of pressure, temperature, and heat release rate when the detonation wave enters the water cloud. In Fig. 10, the detonation wave travels before 0.6 ms, and after that detonation extinction is observed. Specifically, the RF and SF are decoupled, characterized by their increased distance $\Delta_x$. For instance, from 1.2 to 2.5 ms, $\Delta_x$ increases from 0.138 to 0.548 m. Ultimately, the RF trails behind the leading SF as a premixed flame in the shocked $H_2$/$O_2$/Ar mixture. The detonation extinction is also accompanied by quickly reduced peak of pressure, temperature, and heat release rate in Fig. 10. Likewise, the unsteady process of the third mode is visualized in Fig. 11. Different from the results in Fig. 10, detonation extinction occurs immediately when the detonation wave enters the two-phase area. Meanwhile, the reaction front is much weaker (with maximum HRR only about $10^4$ J/m$^3$/s at 3.5 ms) than that in Fig. 10. This is due to higher water mass loading, leading to more energy and momentum extraction from the gas phase by the evaporating droplets and thereby weaker chemical reactions.

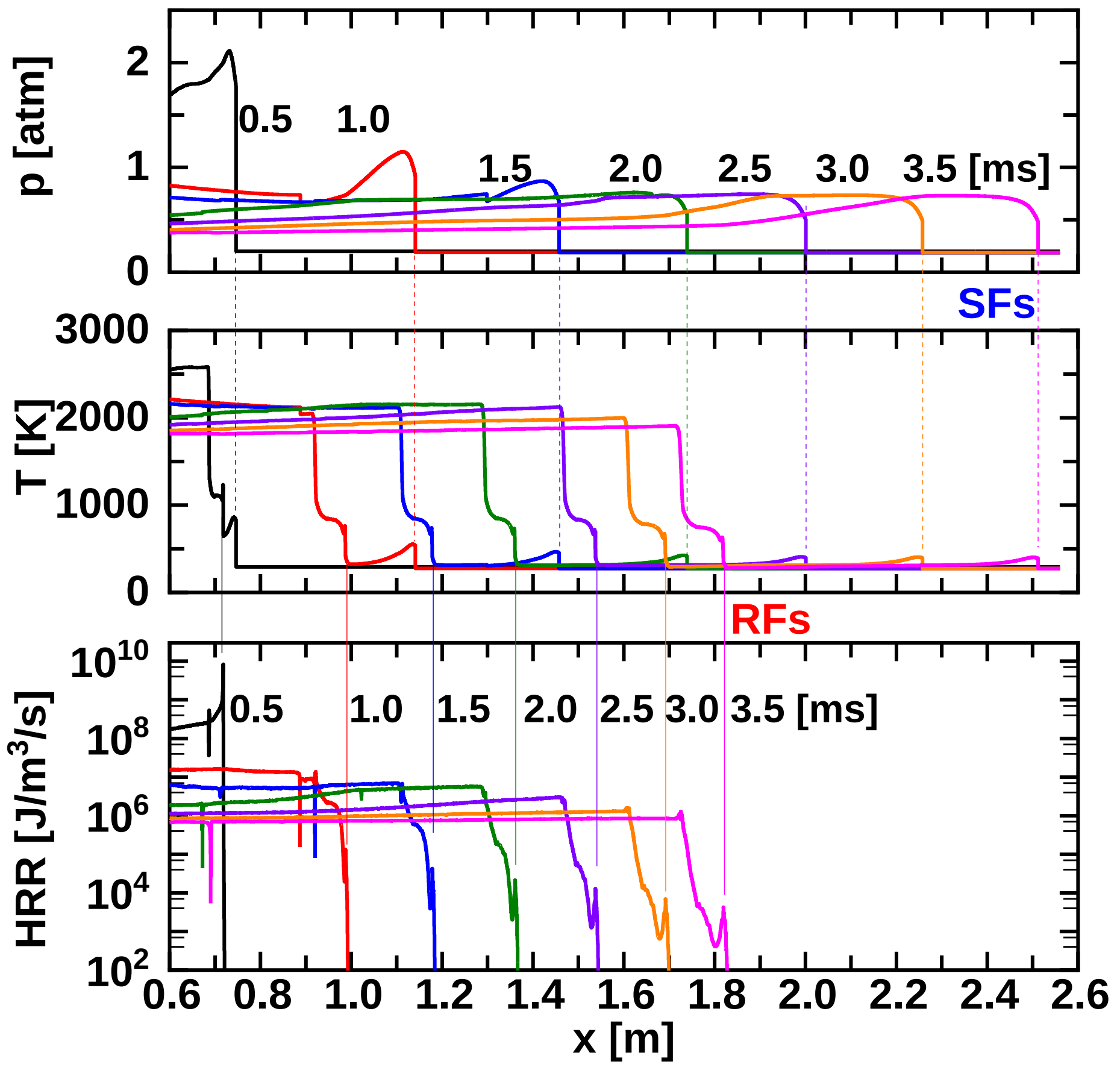

Figure 11 Spatial evolution of pressure, gas temperature, and heat release rate. Dashed lines: SF or RF. $d_d^0$ = 5 μm and $z$ = 0.25.

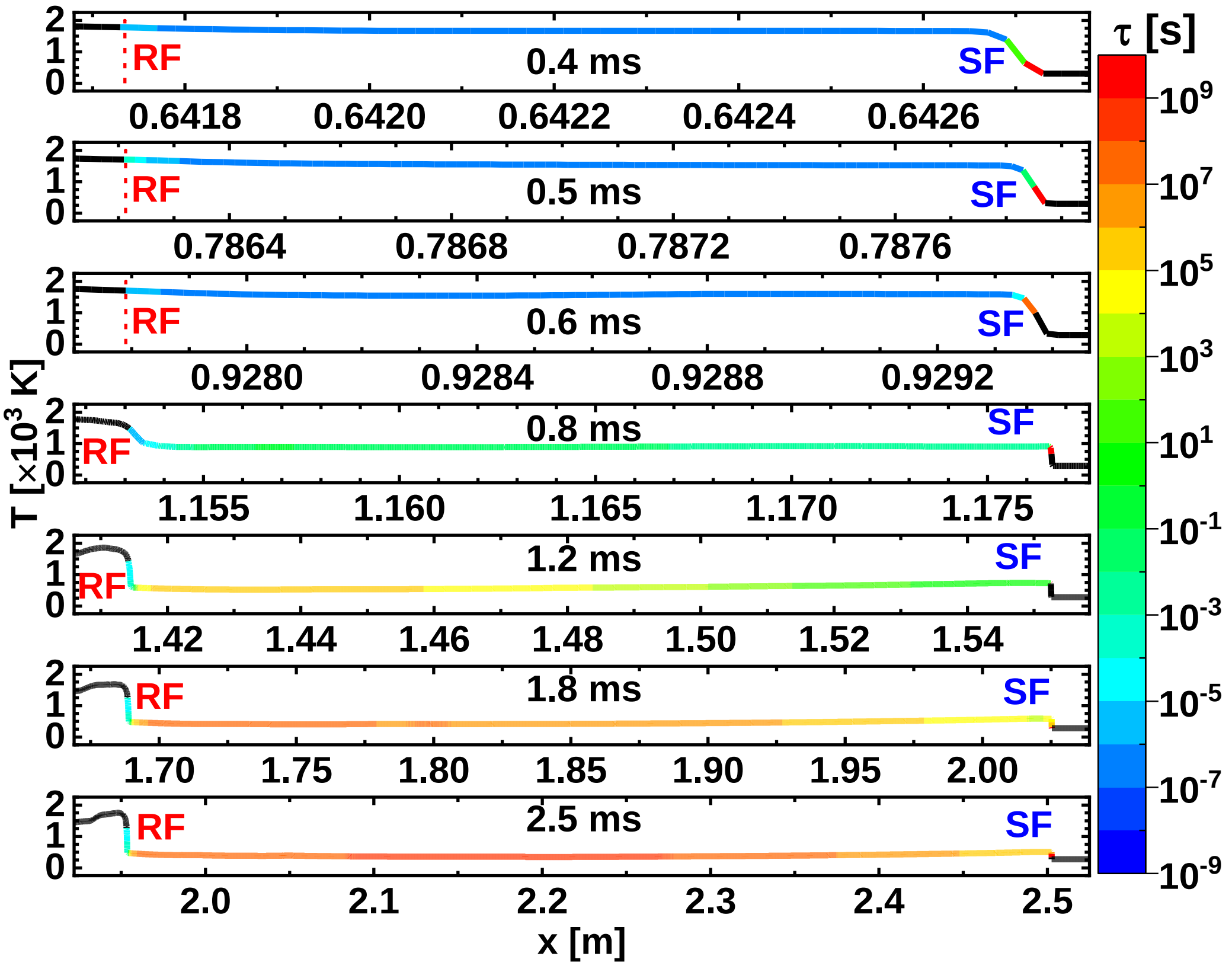

Figure 12 Temporal evolutions of chemical timescale between the RF and SF. $d_d^0$ = 5 μm and $z$ = 0.09.

Figures 12 and 13 demonstrate the distributions of chemical timescale τ in the induction zone of the above two cases, which are colored along the gas temperature profiles. They correspond to 0.4 to 2.5 ms, and 0.5 to 3.5 ms, respectively. $\tau$ is the reciprocal of the real part of the eigenvalues, calculated from the chemical explosive mode analysis [26,28]. Only the chemically explosive mixtures (with positive real part of the eigenvalues from the chemical Jacobian matrix) are colored. At 0.4−0.6 ms for detonative combustion in Fig. 12, the chemical timescale is short, in the order of $10^{-7}$ — $10^{-5}$ s, indicating that the chemical reactions are still strong in the entire induction zone. However, from 0.4 to 0.6 ms, the induction zone length $\Delta_x$ increases from 1 to 1.6 mm, due to the heat absorption and/or shock attenuation by the evaporating water droplets [63–66]. Subsequently, as the RF gradually decouples from the SF after 0.8 ms, and the detonation wave is degraded into a blast wave. Meanwhile, the induction zone length is further increased to 0.6 m at 2.5 ms. Also, the chemical timescale is dramatically increased to above $10^{-3}$ s. The intrinsic chemical explosion propensity of the shocked mixture between RF and SF is significantly reduced. This thereby prevents the possibility of local autoignition in the shocked gas, as observed in pulsating detonations (see Figs. 6 and 7).

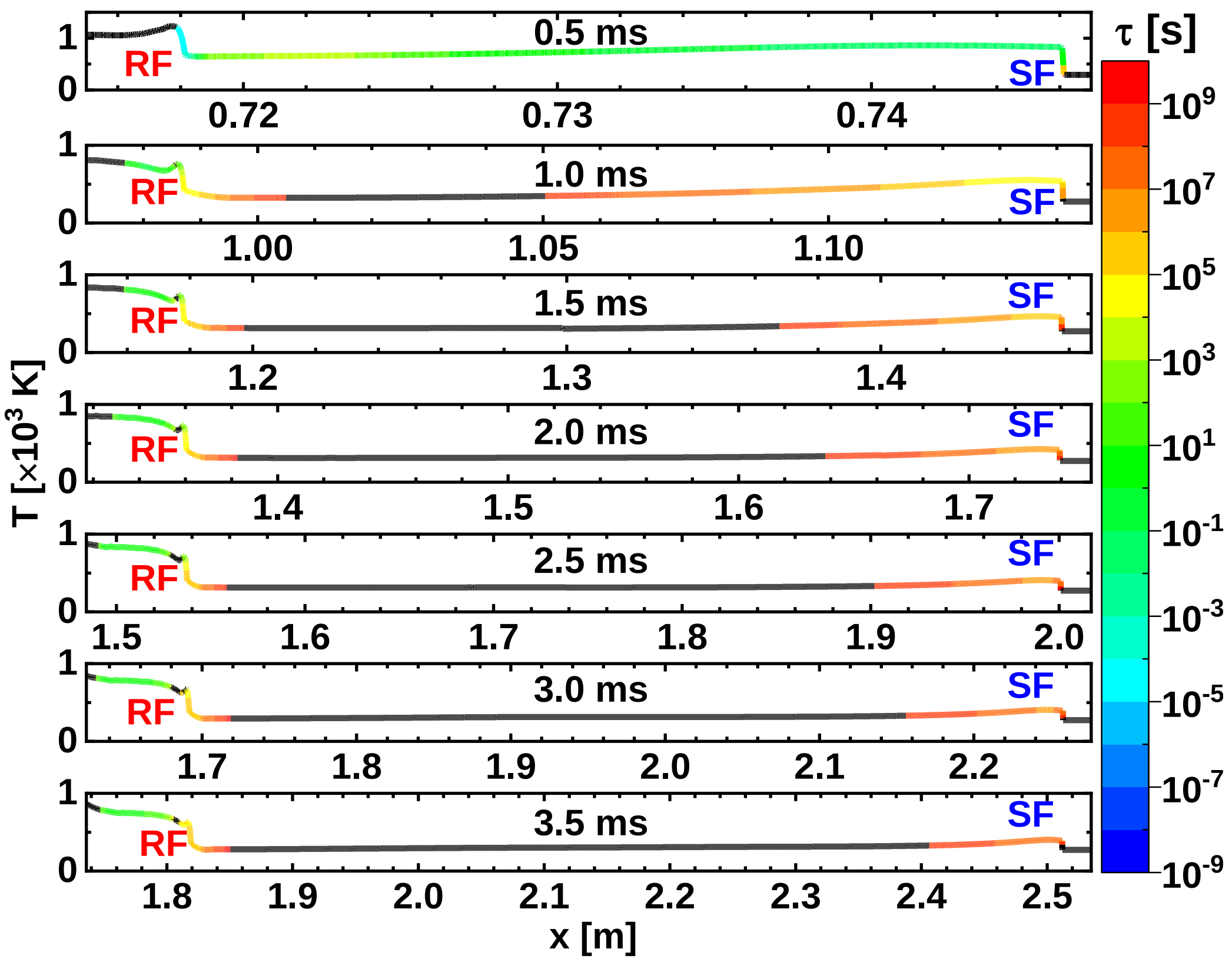

Figure 13 Temporal evolutions of chemical timescale between the RF and SF. $d_d^0$ = 5 μm and $z$ = 0.25.

Figure 13 shows the evolutions of the chemical timescale from the third mode with higher mass loading. They are generally similar to the results in Fig. 12, but there are still striking differences. Firstly, at 0.5 ms, when the DW just enters the water cloud, the chemical timescale is much longer than the counterpart instant (e.g., 0.4 ms) in Fig. 12, indicating the stronger effects of water droplets on the chemical reactions behind the leading shock. Secondly, since 1.0 ms, the shocked mixture in the middle of the induction zone becomes not chemically explosive (see the black segments in Fig. 13) and hence the mixtures with explosion propensity become discontinuous. Specifically, the explosive mixtures are only present immediately behind the leading SF and before the RF. These may be because the relatively high local temperature and/or radical diffusion from the RF. Meanwhile, as the distance between SF and RF increases from 1.0 ms to 3.5 ms, the extent of the explosive mixtures in the induction zone gradually shrinks, which is particularly obvious behind the SF. In addition, in the third mode, the RF is much weaker than the counterpart in Fig. 12. Therefore, even behind the RF, the mixture has chemical explosion propensity. Compared to the second mode, the present one is featured by a more effective inhibition of hydrogen detonation by evaporating water sprays: not only do they quench the incident detonation wave, but also considerably reduce the decoupled blast wave intensity and RF reactivity. From practical explosion hazard prevention, the latter is also of utmost importance, due to possible structural destruction by the blast wave [67] and deflagration-to-detonation transition for new explosion accidents [68]. Critical mass loadings to achieve the third mode and whether the effectiveness would be continuously enhanced through gradually increasing the sprayed water loading will be further discussed in Section 4.4.

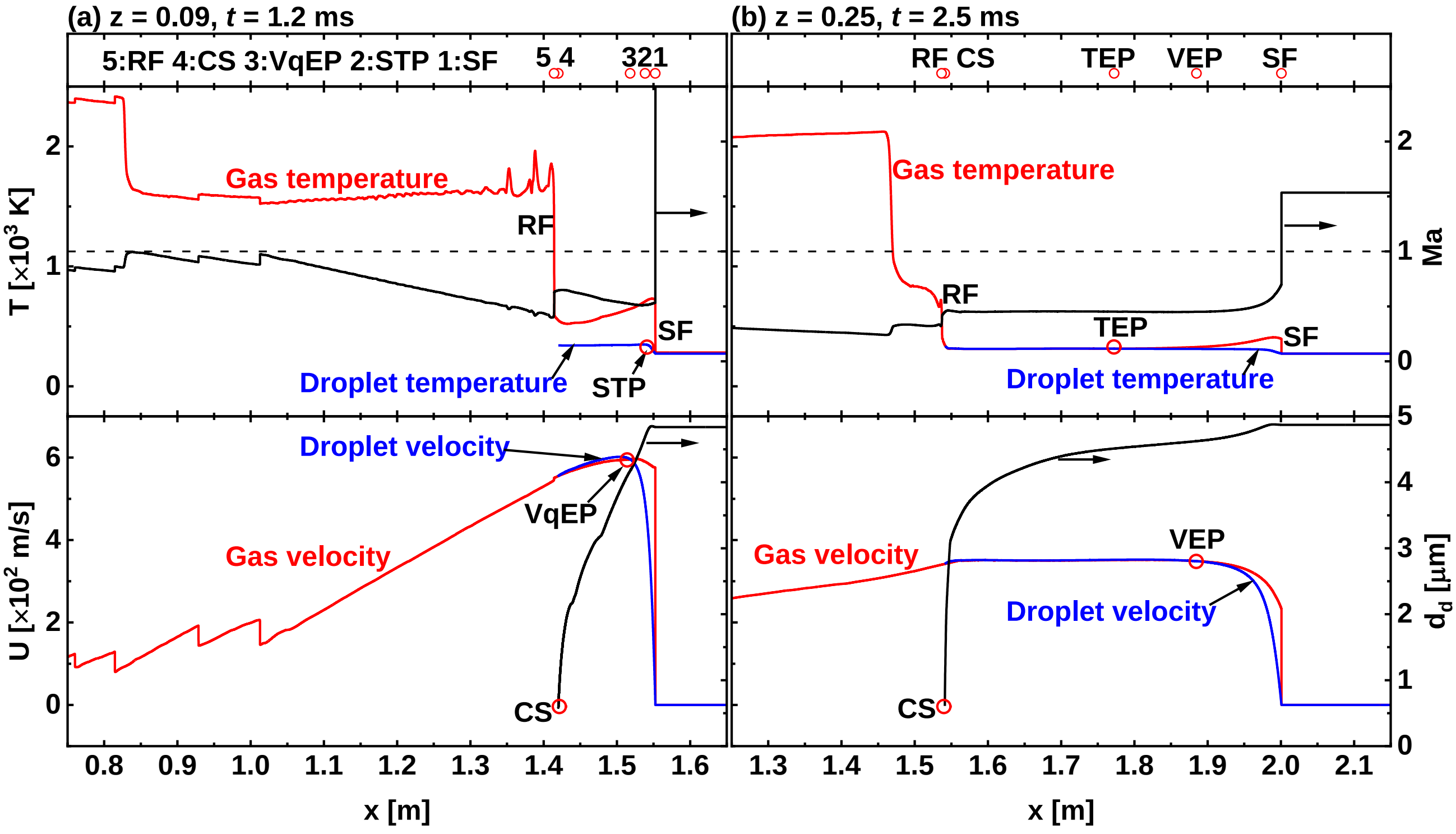

Figure 14 Distributions of temperature, velocity, shock-frame Mach number and droplet diameter: (a) $z = 0.09$ and (b) $z = 0.25$. $d_d^0 = 5$ μm. CS: two-phase contact surface; STP: saturation temperature point; TEP: temperature equilibrium point; VEP: velocity equilibrium point; VqEP: velocity quasi-equilibrium point.

Figure 14 depicts the spatial profiles of temperature, velocity, shock-frame Mach number, and droplet size in the foregoing detonation extinction cases. They respectively correspond to the instants of 1.2 and 2.5 ms. Through them, the hydrodynamic structure of spray detonations can be identified, similar to Fig. 10. They are projected to the *x*-axis and shown at the top of the figures. It is observed from Fig. 14(a) that the relative positions of the SF, RF, STP, VqEP, and two-phase CS have changed, compared to the detonation results in Fig. 8. Specifically, the RF is well behind all other locations, and the shocked flow velocity relative to the leading SF is subsonic (hence absent SP). From Fig. 14(b), one can see that these locations have evolved into the following order, i.e., SF, velocity equilibrium point (VEP), temperature equilibrium point (TEP), CS and RF. Different from the situations in Figs. 14(a) and 8, the kinematic and thermal equilibria between the shocked gas and water droplets are achieved. This is because the lengthened distance between SF and RF and therefore the momentum and thermal relaxation timescales can be reached.

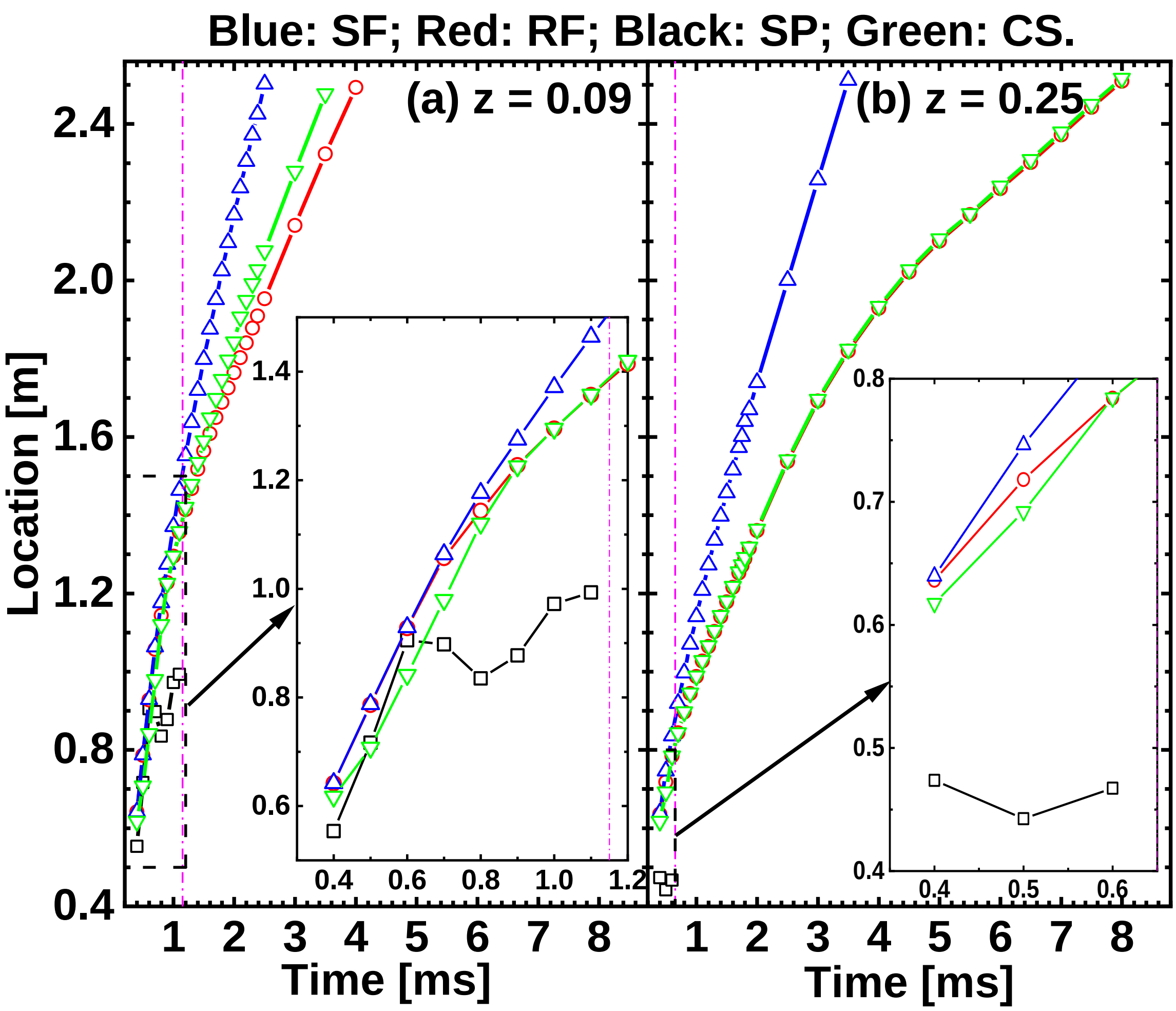

Figure 15 Time evolutions of characteristic locations in the course of detonation extinction: (a) $z$ = 0.09 and (b) z = 0.25. $d_d^0$ = 5 μm. Dash-dotted lines demarcate homogeneous and heterogeneous RF.

The time evolutions of spray detonation flow structure in the course of detonation extinction by water sprays is shown in Fig. 15. When the water loading is 0.09, it is observed from Fig. 15(a) that the distance between the SF and RF gradually increases. Before 1.2 ms (dashed -dotted line), the two-phase contact surface is well behind the SF and RF (see the inset of the Fig. 15a). Therefore, the mixture around the RF is *heterogeneous* (gas and droplets). After that, due to increased induction length, the droplets complete the evaporation well ahead of the reaction zone (i.e., $x_{CS} > x_{RF}$), as seen from Fig. 15(a). This indicates that after detonation extinction, the mixture before the RF is *homogeneous* (gas only) and the RF evolves as a gaseous premixed flame in shocked $H_2/O_2/Ar/H_2O$ mixture. Water vapour addition as a diluent and latent heat absorption jointly reduce the chemical reactivity. This can also be observed when the water mass loading is 0.25 in Fig. 15(b). However, because of high water loading, the distance between the two-phase contact surface and RF is smaller in Fig. 15(b). The homogeneous and heterogeneous flame propagation in water sprays is found to have different dynamic behaviours, based on the recent theoretical analysis for laminar flames by Zhuang and Zhang [69,70]. Further studies are merited, from both fundamental and practical hazard prevention perspectives, to unveil the interactions between the residual RF and dispersed droplets after detonation extinction. In addition, from Fig. 15(b), the RF and CS have similar propagation speeds, e.g., about 145 m/s after 4 ms. For both cases, after detonation extinction, the shock-frame sonic points (see the squares in Fig. 15) generally move downstream after the detonation encroaches the water clouds. Beyond about 1.1 ms and 0.7 ms respectively in two cases (see their insets), the shocked gas become subsonic in the shock frame.

### 4.4 Diagram of detonation propagation and extinction in water sprays

The unsteady detonation phenomena, including detonation pulsation propagation and extinction, are further generalized through considering different initial droplet diameters $d_d^0$ and mass loadings $z$. The results are summarized in the diagram of droplet size and mass loading in Fig. 16. The dashed line is the critical condition for extinction (including the second and third modes) of the incident $H_2/O_2/Ar$ detonations in the water sprays. Apparently, when the initial droplet size is reduced, the critical mass loading for detonation extinction, $z_{ext}$, decreases considerably. For instance, $z_{ext}$ = 0.08 for 5 μm, whilst it is increased to 0.375 for 15 μm. This is reasonable because smaller droplets have fast evaporation rate and larger specific droplet surface area and hence are more effective in inhibiting the detonation wave. Meanwhile, when the water loading is sufficiently large, e.g., $z > 0.2$, detonation extinction always occurs (except the case of $z$ = 0.025 and $d_d^0$ = 15 μm), for all the considered droplet size. Moreover, when z < 0.07, the detonation wave can always successfully cross the water cloud in a galloping fashion, regardless of droplet size. However, when $z$ = 0.09 and 0.1, the droplet diameter effect becomes pronounced: for these two loadings, finer droplets with $d_d^0 \leq 5$ μm are necessary to quench the incident detonations.

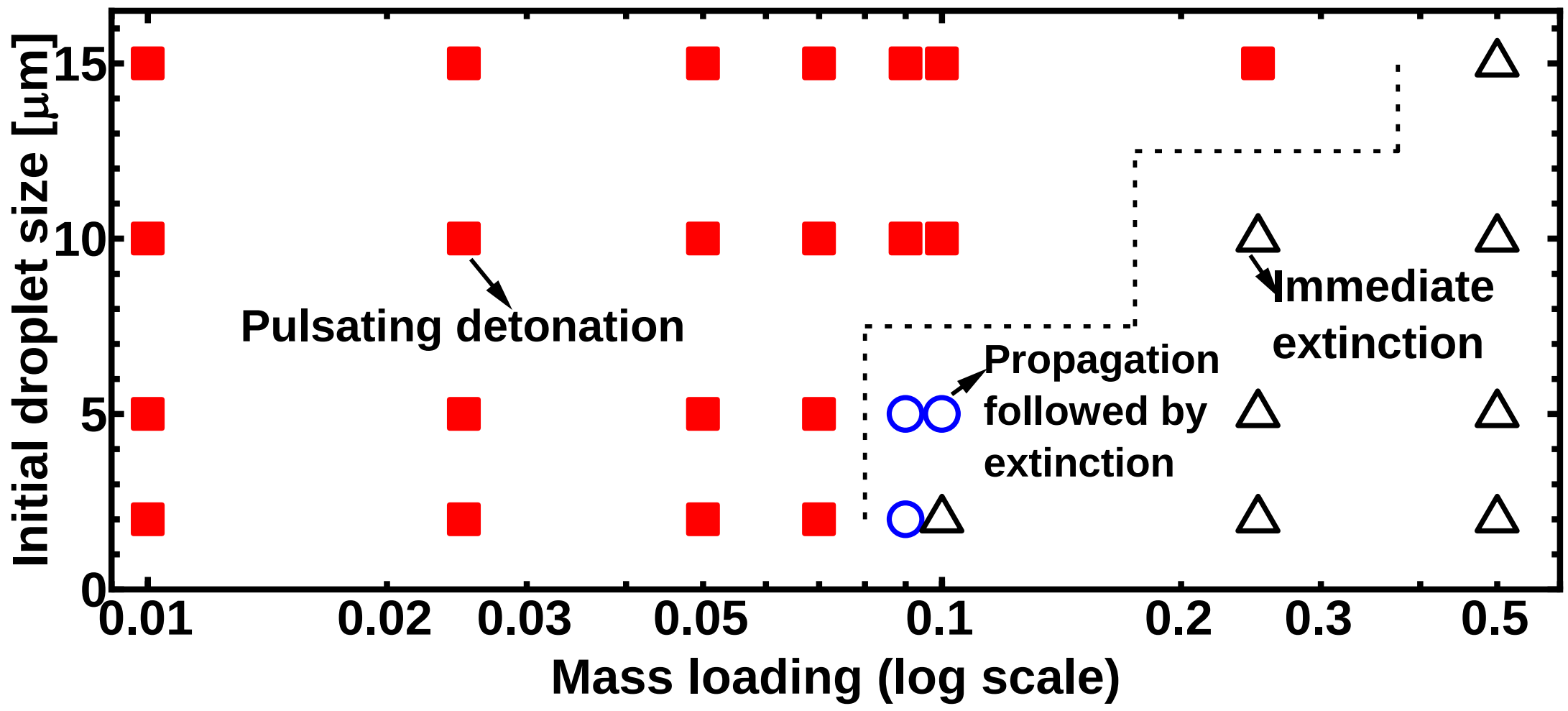

Figure 16 Diagram of detonation propagation and extinction in fine water sprays. Squares: detonation propagation; triangles: immediate extinction; circles: propagation followed by extinction.

To measure the effectiveness of various water droplets in detonation inhibition, extinction time and distance in the detonation failure cases are shown in Fig. 17. In our study, they are respectively defined as the duration and distance when the detonation travel in the water sprays and the leading shock speed is reduced to half of the C—J velocity of the dry mixture $D_{CJ}$. For a fixed water spray diameter, both extinction time and distance decrease monotonically with mass loading. Meanwhile, when the mass loading is fixed, the extinction time / distance monotonically increases with the droplet diameter. Interestingly, when the mass loading is increased beyond 0.5, the extinction time / distance corresponding to three diameters are very close. Therefore, continuously increasing the water mass loading does not proportionally increase the effectiveness of fine water sprays for detonation extinction; instead, the extinction time and distance would approach the asymptotic values, i.e., 0.1 m and 0.07 ms. This is essentially limited by the finite timescale of two-phase coupling between gas and droplet phases, such as heat transfer and droplet evaporation. For practical implementations of explosion prevention, this result implies that: (1) when the water loading is sufficiently high, the effectiveness in detonation inhibition has weak dependence on sprayed droplet sizes; and (2) these asymptotic values can be important reference for designing water curtain dimensions.

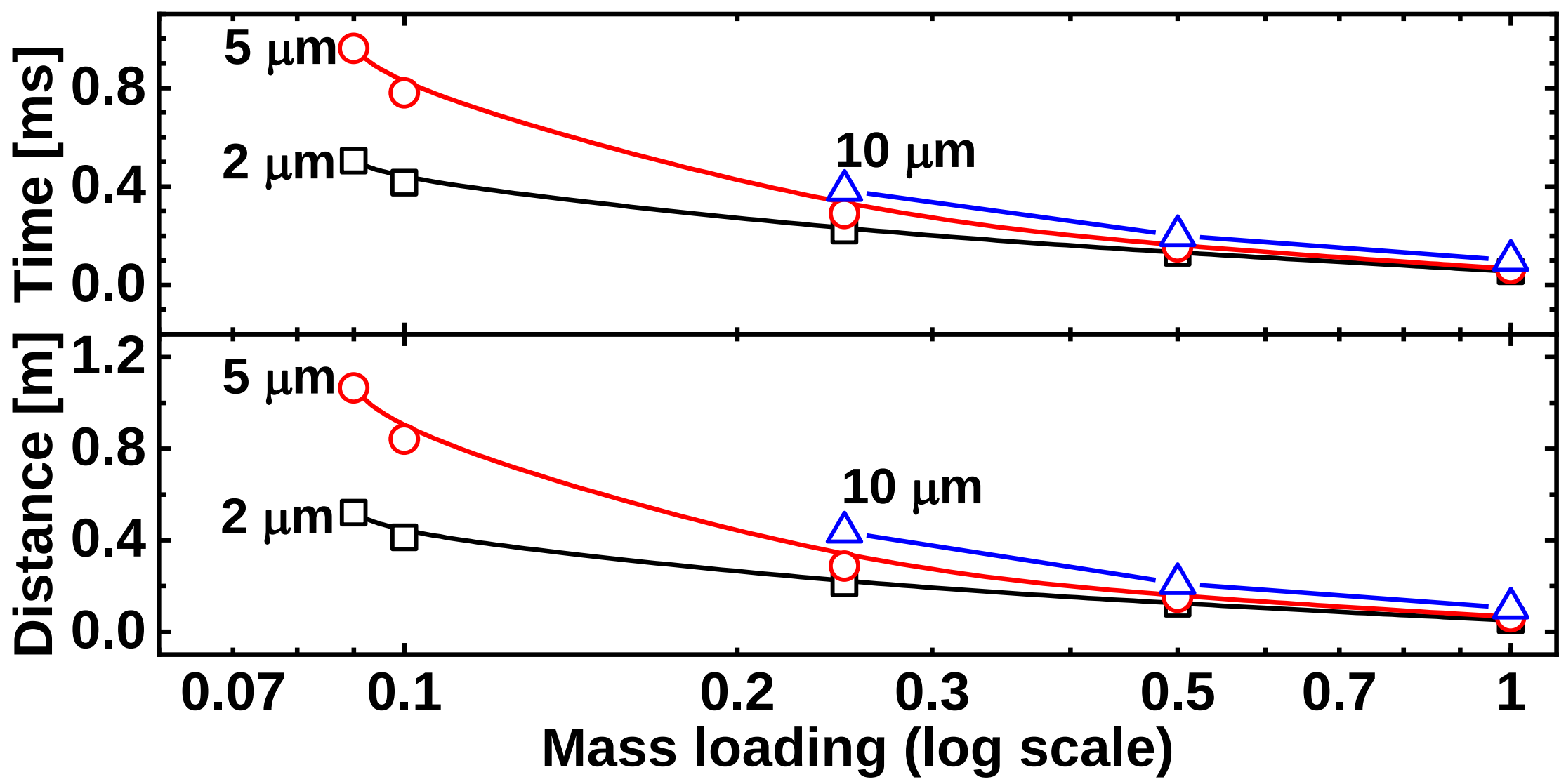

Figure 17 Extinction time and distance as functions of water mass loading and initial droplet size.

Pulsating detonations are seen in all the stable cases in Fig. 16, and their average leading shock speed is shown in Fig. 18. It is observed that the detonation speeds in two-phase medium are consistently lower than that in water-free case ($D/D_{CJ}$ < 1). One can see from Fig. 18(a) that for a fixed water mass loading, the shock speed is higher when the droplet size is higher. This is reasonable because the larger diameter corresponds to a smaller specific surface area, and therefore slower two-phase exchange about mass, momentum, and energy. However, an opposite tendency is seen for fine droplets and a minimum speed exists around $d_d^0$ = 5 μm. When the droplet diameter is smaller than 5 μm, such as $d_d^0$ = 2 μm, the magnitudes of interphase transfer (e.g., energy and momentum absorption from the gas) are lower than those of the droplets with 5 μm. This can be confirmed in Section C of Supplementary Document. Moreover, one can see from Fig. 18(b), the average speeds monotonically decrease with mass loading for a fixed droplet size, because transferred heat and momentum become greater. In addition, the effects of droplet diameter and mass loading on pulsation frequency are shown in Fig. 19. Generally, the frequency varies between 2.6×10$^4$ and 3.1×10$^4$ Hz, except the two case near the extinction critical condition. For all the droplet sizes in Fig. 19, the frequency decreases when the mass loading becomes larger. This is because the RF−SF coupling is weakened by stronger interphase exchange, indicating that the RF and SF need longer time to complete one pulsation cycle. For the same mass loading (such as 0.05), the pulsating frequency shows non-monotonic evolution with initial droplet size. Like the results of shock speed in Fig. 18, the case with $d_d^0$ = 5 μm has the smallest frequency.

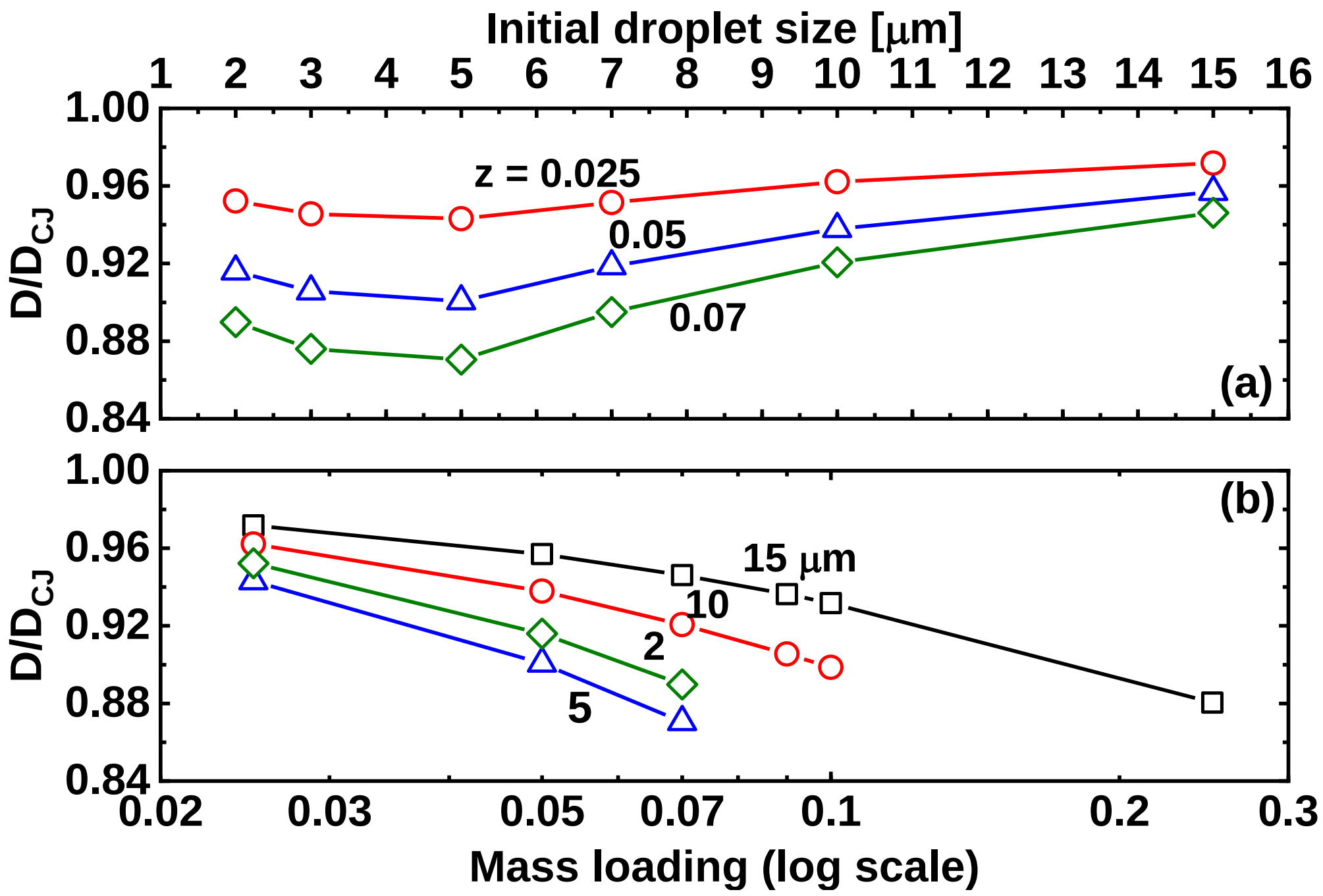

Figure 18 Change of leading shock propagation speed with (a) droplet diameter and (b) mass loading. $D_{CJ}$: C—J speed of water-free $H_2/O_2/Ar$ mixture.

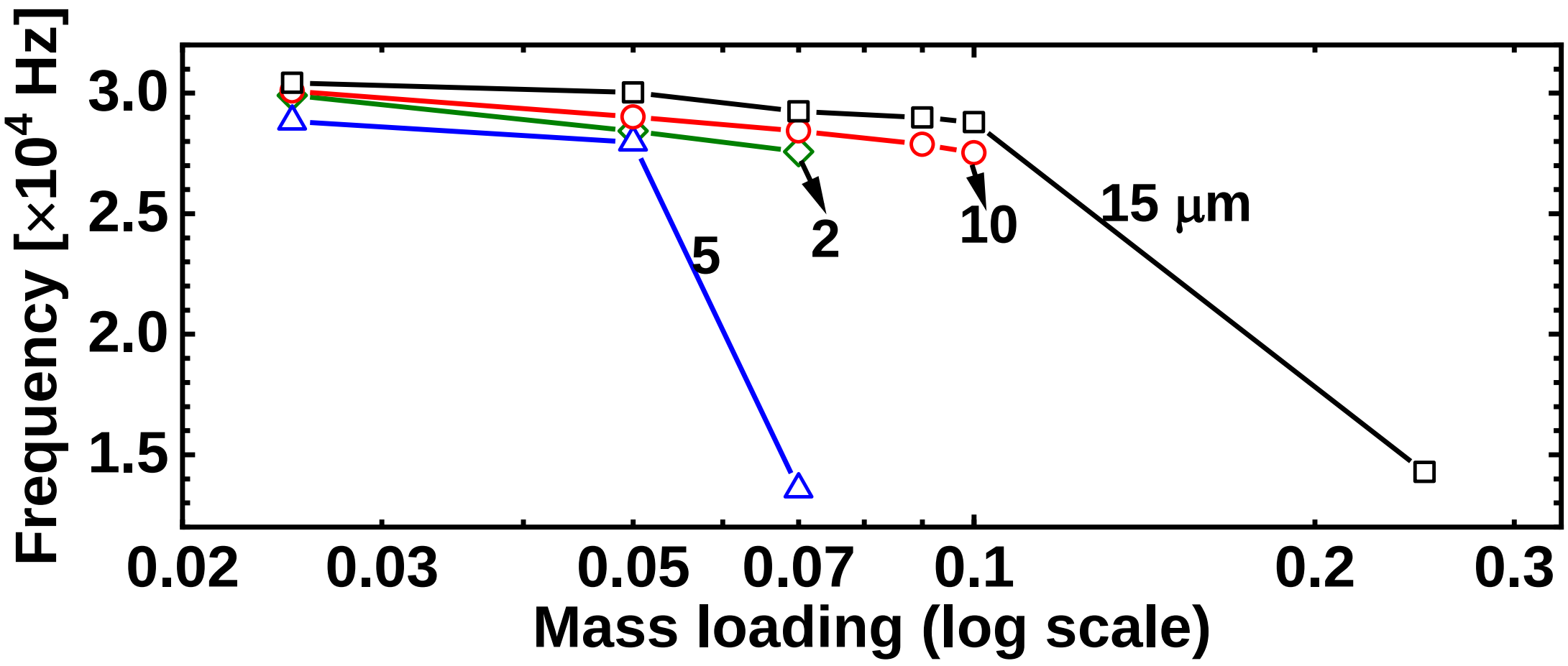

Figure 19 Change of pulsation frequency with mass loading and droplet size.

## 5. Conclusions

Pulsating propagation and extinction of stoichiometric hydrogen/oxygen/argon detonations in water mists are studied using Eulerian−Lagrangian approach with two-way gas−liquid coupling. Detailed chemical mechanism is used for gaseous detonation, and ultrafine mono-sized water droplets are considered, covering a range of droplet size and mass loading.

Three detonation propagation modes are observed: (1) pulsating propagation, (2) extinction after a finite propagation, and (3) immediate extinction. For pulsating detonation, the propagation speeds of RF and SF and their distance is found to have periodic changes. Within one cycle, multiple pressure waves are generated from the RF, which overtake and intensify the leading shock. An autoigniting spot arises in the shocked gas, which is confirmed from the eigenvalue evolutions of the chemical Jacobian matrix in chemical explosive mode analysis. The evolutions of the characteristic locations in the flow structure of spray detonations, including SF, RF, shock-frame sonic point, and two-phase contact surface are also analyzed. The results show that their relative positions remain unchanged within one pulsating cycle, but their distances are shown to have periodic variations.

The unsteady behaviors of detonation extinction are also studied. The results reveal that the decoupling of RF and SF is accompanied by considerably increased chemical timescale of the shocked mixture, and part of the mixture behind SF is even chemically non-explosive due to relatively low temperature. The relative positions of the characteristic locations in spray detonations also change significantly. The reaction front trails behind all other characteristic locations, and the post-shock flow is subsonic in the shock frame. The results also demonstrate that gas-only mixtures exist before the RF due to the complete evaporation of the droplets, and henc the RF behaves like a gaseous premixed flame in $H_2/O_2/Ar/H_2O$ mixture.

The effects of droplet properties on unsteady detonation propagation are summarized in a diagram of droplet size versus mass loading. It shows that the critical mass loading for detonation extinction decreases when the droplet size becomes smaller. Moreover, the detonation extinction distance and time decrease with increased water loading and reduced droplet size. Asymptotic values of extinction time and distance exist (0.1 m and 0.07 ms, respectively) when the water loading is beyond 0.5. Moreover, the frequency and average speed of pulsating detonation is shown to have non-monotonic dependence on the droplet size but monotonically decreases with water mass loading.

**Acknowledgements**

This work used the computational resources of ASPIRE 1 Cluster in The National Supercomputing Centre, Singapore (https://www.nscc.sg). YX is supported by the NUS Research Scholarship. HZ is supported by National Research Foundation in Singapore (R-265-000-A57-592). Professor Zhuyin Ren and Dr Wantong Wu at Tsinghua University are thanked for sharing the CEMA subroutines. Professor Michael P. Burke from Columbia University is also appreciated for helpful discussions about the chemical mechanism.